\documentclass{aa}  

\usepackage{graphicx}
\usepackage{txfonts}
\usepackage{enumerate}

\usepackage{amsmath}
\usepackage{amssymb}

\usepackage{rotating}

\usepackage{soul}
\usepackage[usenames,dvipsnames]{color}

\usepackage{url}

\definecolor{vertdc1}{RGB}{20,89,33}
\definecolor{blood}{RGB}{193,41,41}
\definecolor{viol}{RGB}{109,10,186}
\definecolor{dgreen}{RGB}{9,95,29}
\definecolor{DarkGreen}{RGB}{9,95,29}
\definecolor{dorange}{RGB}{197,69,6}

\definecolor{CNRSBlue}{RGB}{26,48,81}
\definecolor{CNRSLightBlue}{RGB}{10,141,167}
\definecolor{DarkRed}{RGB}{212,0,0}
\definecolor{DarkRed2}{RGB}{150,0,0}
\definecolor{Violet}{RGB}{122,47,214}

\usepackage[switch,columnwise]{lineno}

\usepackage{hyperref}
\hypersetup{
colorlinks=true,
breaklinks=true,
urlcolor= RedViolet,
citecolor=blue,
linkcolor=blue,
breaklinks=true
}
\usepackage{breakurl}

\usepackage[framemethod=tikz]{mdframed}

\let\FMequation\equation
\let\endFMequation\endequation

\renewenvironment{equation}{\begin{linenomath*}\FMequation}{\endFMequation\end{linenomath*}}
\begin{document} 

%\linenumbers

\title{Vertical compositional variations of liquid hydrocarbons in Titan's alkanofers}

\author{Daniel Cordier\inst{1}\fnmsep\thanks{\email{daniel.cordier@univ-reims.fr}}
        \and
        David~A.~Bonhommeau\inst{1}
        \and
        Tuan~H.~Vu\inst{2}
        \and
        Mathieu~Choukroun\inst{2}
        \and
        Fernando Garc\'{i}a-S\'{a}nchez\inst{3}
        }

   \institute{Universit\'{e} de Reims Champagne Ardenne, CNRS, GSMA UMR CNRS 7331, 51097 Reims, France.\\
         \and NASA Jet Propulsion Laboratory, California Institute of Technology, Pasadena, CA 91109, USA.\\
         \and Engineering Management of Additional Recovery,
                 Mexican Petroleum Institute,
                 Eje Central L\'{a}zaro C\'{a}rdenas Norte 152, 
                 07730 Mexico City, Mexico.}

   \date{Received March 12, 2021; accepted June 22, 2021}

% \abstract{}{}{}{}{} 
% 5 {} token are mandatory
 
  \abstract
   {According to clues left by the Cassini mission, Titan, one of the two Solar System bodies with a hydrologic cycle, 
   may harbor liquid hydrocarbon-based analogs of our terrestrial aquifers, referred to as ``alkanofers''.}
   {On the Earth, petroleum and natural gas reservoirs show a vertical gradient in chemical composition, established over 
  geological timescales. In this work, we aim to investigate the conditions under which Titan's processes could lead to similar situations.}
   {We built numerical models including barodiffusion and thermodiffusion (Soret's effect) in 
  N$_2$+CH$_4$+C$_2$H$_6$ liquid mixtures, which are relevant for Titan's possible alkanofers. Our main assumption is the existence of reservoirs 
  of liquids trapped in a porous matrix with low permeability.}
   {Due to the small size of the molecule, nitrogen seems to be more sensitive to gravity than ethane, 
  even if the latter has a slightly larger mass. This behavior, noticed for an isothermal crust, is reinforced by the presence of a geothermal
  gradient. Vertical composition gradients, formed over timescales of between a fraction of a mega-year to several tens of mega-years, are 
  not influenced by molecular diffusion coefficients. We find that ethane does not accumulate at the bottom of the alkanofers under diffusion, 
  leaving the question of why ethane is not observed on Titan's surface unresolved.
  If the alkanofer liquid was in contact with water-ice, we checked that N$_2$ did 
  not, in general, impede the clathration of C$_2$H$_6$, except in some layers. Interestingly, we found that noble gases could easily accumulate
  at the bottom of an alkanofer.}
  {}


\maketitle
%

%%%%%%%%%%%%%%%%%%%%%%%%%%%%%%%%%%%%%%%%%%%%%%%%%%%%%%%%%%%%%%%%%%%%%%%%%%%%%%%%%%%%%%%%%%%%%%%%%%%%%%%%%%%%%%%%%%%%%%%%%%%%%%%%%%%%
\section{\label{intro}Introduction}

% Quelques généralités sur Titan :
  Since the detection of its thick atmosphere by G.~Kuiper \citep{kuiper_1944}, Titan, the main satellite of Saturn, has been the subject
of many studies \citep[see, for instance, the monograph edited by][]{mullerwodarg_etal_2014} and targeted by two major space missions: Voyager 
and Cassini, while an in situ exploration by the revolutionary rotorcraft Dragonfly \citep[see, for example,][]{turtle_etal_2020} 
is in preparation. 
This unfailing interest led the planetary science community to numerous important discoveries. While expected for decades 
\citep{tyler_etal_1981,owen_1982,sagan_dermott_1982,flasar_1983,eshleman_etal_1983}, lakes and seas of liquid hydrocarbons were discovered 
by Cassini's instruments in Titan's polar regions \citep{stofan_etal_2007,stephan_etal_2010}. These observations have opened 
the door to a unique case of ``exo-oceanography'', connected to an ``exo-hydrology'' (or ``alkanology'') for which liquid methane is the 
main working fluid.\\
  In addition to seas and lakes, dry fluvial channels have been detected \citep{legall_etal_2010,coutelier_etal_2021} and evidence of the presence 
of evaporite deposits \citep{barnes_etal_2011,cordier_etal_2013b,macKenzie_etal_2014,cordier_etal_2016b} and underground alkanofers has 
been found \citep{corlies_etal_2017,hayes_etal_2017}. These latter geological structures were proposed to be the Titanian analogs 
for terrestrial aquifers. While a porous water-ice matrix plays the role of Earth's porous rocks, the liquid contained in the pores 
is a mixture of methane and ethane, complemented by some amount of dissolved nitrogen \citep{cordier_etal_2017a}.\\
The fast destruction rate of atmospheric methane predicted by pre-Cassini works \citep{flasar_etal_1981,yung_etal_1984} lead scientists 
to hypothesize the existence of a subsurface source of methane. 
One potential endogenic source of methane could be the complete or partial dissociation of a clathrate hydrate reservoir 
\cite[e.g.,][]{lunine_stevenson_1987,tobie_etal_2006,choukroun_etal_2010,munoz-iglesias_etal_2018,petuya_etal_2020}
 In addition,  the possible presence of liquid reservoirs, 
more or less deeply buried under Titan's surface, has been proposed. For instance, \cite{griffith_etal_2012} interpreted dark patches 
observed in equatorial regions as an emergence of liquid. A subsurface alkanofer is also mentioned by \cite{sotin_etal_2012} in their discussion 
of Titan's organic cycle. \cite{mousis_etal_2014} focused their study on the possible interplay between liquids and an icy matrix 
at least partially made of clathrates. In order to explain the cloud distribution observed by Cassini instruments, \cite{turtle_etal_2018} 
proposed the existence of a widespread polar subsurface methane reservoir. In the same vein, \cite{macKenzie_etal_2019} mentioned
a drainage effect in a subsurface reservoir for interpretation of surface changes.
However, the best indirect evidence of alkanofer existence to date was found by \cite{hayes_etal_2017}. 
Indeed, the Cassini radar instrument has permitted altimetric measurements of polar regions. In this way, 
\cite{hayes_etal_2017} found that several maria have their surfaces along the same gravity equipotential level, suggesting the existence of 
some subterranean connectivity.\\
   The study of possible Titan alkanofers is stimulating in many respects. First of all, the methane potentially stored in these reservoirs may 
participate significantly in the global carbon cycle of Titan \citep{horvath_etal_2016,faulk_etal_2020}. Secondly, the lack of ethane 
on the surface of Titan is a long-standing problem \citep{mousis_etal_2016,gilliam_lerman_2016}; a response to this question could consist 
of a chemical vertical stratification of subsurface liquids. Since ethane (molecular weight: $30$~g~mol$^{-1}$) is heavier 
than nitrogen ($28$~g~mol$^{-1}$) and methane ($16$~g~mol$^{-1}$), 
ethane-enriched layers could exist in Titan's deep alkanofers. 
Finally, the nature, the amplitude, and the temporality of exchanges between 
Titan's interior and its atmosphere is a ``cold case'', with an exobiological importance, for which any progress is welcome
\citep{nixon_etal_2018,kalousova_sotin_2020}.\\
   The variation of species in terrestrial hydrocarbon reservoirs is a well-established topic of great industrial interest
\citep[see reviews:][]{Chilingar_etal_2005,obidi_2014,esposito_etal_2017}. In most of the field measurements, a vertical 
variation of chemical composition is detected \citep[see, for example,][]{metcalfe_etal_1988}, with the lightest compounds floating above heavier 
ones. However, surprisingly, the reversed situation is also observed \citep{temeng_etal_1998}. Here, we mainly address the 
question of a similar compositional grading in Titan's crustal hydrocarbon reservoirs. Our approach does not imply large-scale
hydrodynamic models like in recent works \citep{horvath_etal_2016,faulk_etal_2020}, but rather local models focused on species diffusion in
a porous icy matrix with a small hydraulic permeability that impedes convective transport. We propose a transposition of the physics 
of terrestrial oil reservoirs to the Titanian context.\\
   In this article, we first discuss the case of an isothermal system. We subsequently include the effect of the geothermal gradient, and 
we describe our in-depth study of the role of molecular diffusion coefficients. The last section is dedicated to a general discussion about some  
chemo-physical properties of alkanofers.
%
%===================================================================================================================================
\begin{figure*}[!tp]
\begin{center}
\includegraphics[angle=0, width=18 cm]{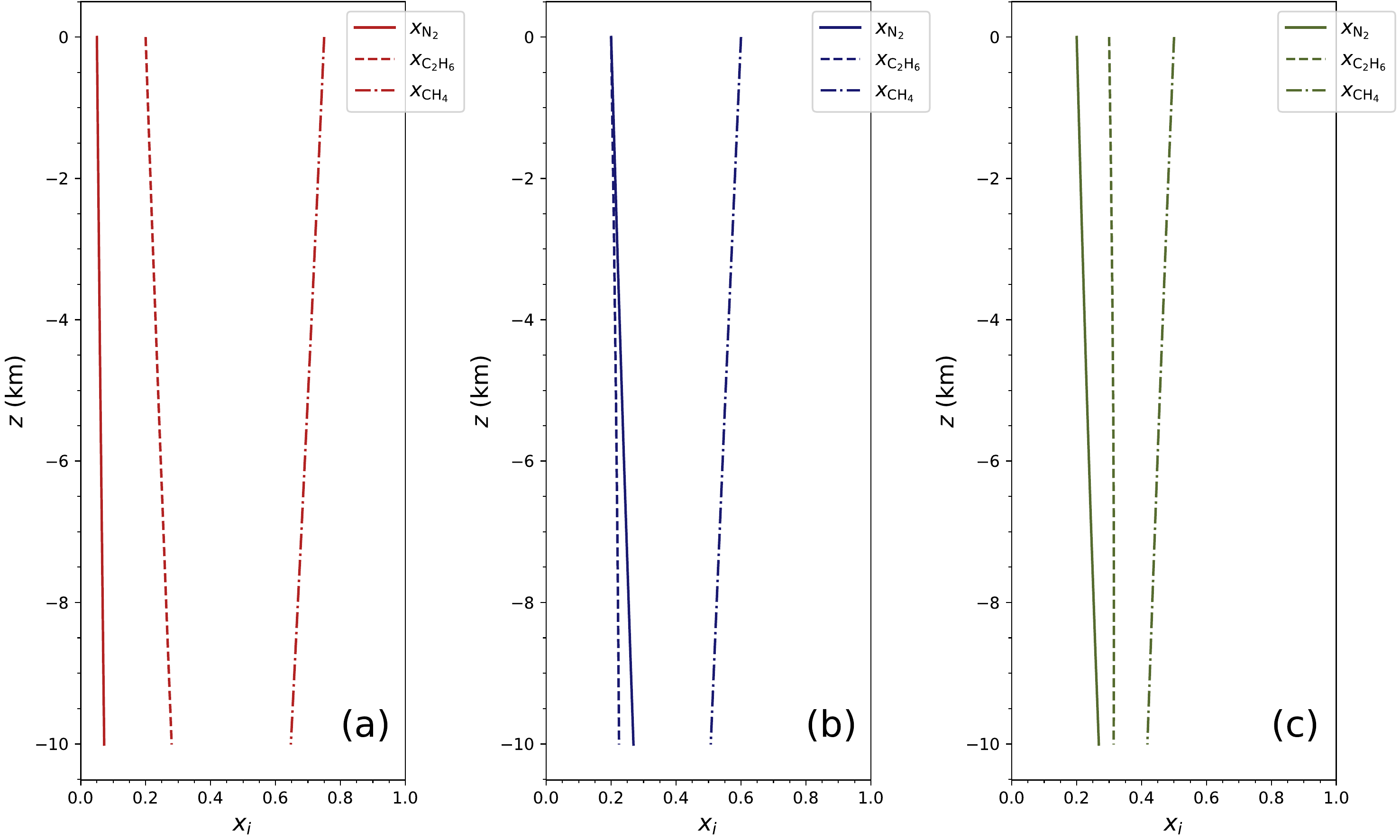}
\caption[]{\label{grad_isotherm_crust}Vertical variations of mole fractions (denoted $x_i$) of the main components (i.e., N$_2$,
           C$_2$H$_6$ and CH$_4$) of Titan's alkanofer liquid. This figure depicts a scenario where the icy crust is assumed to be
           isothermal at the temperature $T_0= 90$~K. The pressure varies between $1.5$~bar at the surface ($z= 0$) and around $130$~bar 
           at the bottom of the simulated system (depth of $z= 10$~km). The three panels correspond to different surface compositions, 
           expressed in mole fractions: 
           (a) N$_2$: $0.05$, C$_2$H$_6$: $0.20$, CH$_4$: $0.75$;
           (b) N$_2$: $0.20$, C$_2$H$_6$: $0.20$, CH$_4$: $0.60$;,           (c) N$_2$: $0.20$, C$_2$H$_6$: $0.30$, CH$_4$: $0.50$. For all computations, the porosity of the water ice matrix was 
           fixed to 5\%, the diffusion theory formalism follows \cite{ghorayeb_firoozabadi_2000}, the water ice equation of state was provided by 
           \cite{feistel_wagner_2006}, while the thermodynamic properties of liquids were computed using PC-SAFT.}
\end{center}
\end{figure*}
%===================================================================================================================================
% 
%
%%%%%%%%%%%%%%%%%%%%%%%%%%%%%%%%%%%%%%%%%%%%%%%%%%%%%%%%%%%%%%%%%%%%%%%%%%%%%%%%%%%%%%%%%%%%%%%%%%%%%%%%%%%%%%%%%%%%%%%%%%%%%%%%%%%%
%% Modèle sans prise en compte du possible gradient thermique dans le sous-sol :
%\clearpage
\section{Case of an isothermal crust}
\label{isothermcrust}

Contrary to situations where a binary mixture is considered, the diffusion processes in 
a ternary mixture can no longer be described by a single equation \citep[similar to Eq.~1 in][]{cordier_etal_2019}. For a 
fluid containing $N$ species, the vectorial diffusion flux $\scriptstyle\overrightarrow{J}$ (kg~m$^{-2}$~s$^{-1}$) has $N-1$ components, and 
can be written \citep{ghorayeb_firoozabadi_2000}, for a 1D system related to a vertical $z$-axis, as
\begin{equation}\label{GeneDiffEqua}
\vec{J} =   -c \, \left(D^{M} \, \frac{\partial\vec{x}}{\partial z} 
                     + D^P \, \frac{\partial P}{\partial z} 
                     + D^T \, \frac{\partial T}{\partial z}\right)
,\end{equation}
with $c$ being the total molar density (mol~m$^{-3}$), $D^{M}$ the $(N-1) \times (N-1)$ molecular diffusion tensor, $D^P$ the $(N-1)\times 1$ 
barodiffusion tensor, and $D^T$ the $(N-1)\times 1$ thermodiffusion tensor. Among these quantities, $P$ is the pressure (in Pa) and $T$ the temperature 
(in K). We also have $\vec{x} = (x_1, x_2, \ldots, x_{N-1})$, with $x_i$ being the mole fraction of species $i$, taken under a condition of normalization.
In the phenomenological Eq.~\ref{GeneDiffEqua}, the three terms in the right hand side correspond to different microscopic transport processes.
The first one represents the well-known molecular fickean diffusion, driven by compositional gradients. The second term corresponds to the diffusion 
induced by pressure variations (also known as barodiffusion), which may lead to gravity segregation. The last term represents thermal diffusion, 
alternatively called the ``Soret effect'' in the case of liquids \citep{soret_1879,esposito_etal_2017}. This effect is linked to the tendency for species, in a nonconvective mixture, to separate themselves under the influence of a temperature gradient.
The generic forms of tensors $D^M$, $D^P,$ and $D^T$ are available in the literature \citep{ghorayeb_firoozabadi_2000}; they specifically depend 
on the fickean diffusion coefficients, on the derivative of fugacities, and on the thermal diffusion coefficients. 
Here, the liquid under consideration involved three species: 
N$_2$, C$_2$H$_6$ and CH$_4$.
In such a case, at equilibrium (i.e., when $\scriptstyle\vec{J}=\vec{0}$ everywhere), the vectorial Eq.~\ref{GeneDiffEqua}
may be reformulated as a set of two (here $N-1=3-1=2$) partial differential equations:
\begin{equation}
D_{11}^M \frac{\partial x_1}{\partial z} + D_{12}^M \frac{\partial x_2}{\partial z} + D_1^P \frac{\partial P}{\partial z} 
                                         + D_1^T \frac{\partial T}{\partial z} = 0
,\end{equation}

\begin{equation}
D_{21}^M \frac{\partial x_1}{\partial z} + D_{22}^M \frac{\partial x_2}{\partial z} + D_2^P \frac{\partial P}{\partial z} 
                                         + D_2^T \frac{\partial T}{\partial z} = 0
,\end{equation}
with the $D_{ij}^M$s representing the elements of the diffusion tensor $D^M$. Similarly, the $D_i^P$s and the $D_i^T$s are respective elements of 
$D^P$ and $D^T$. The indices $i$ and $j$ are related to chemical species; we chose $i=1$ for nitrogen, $i=2$ for ethane, and $i=3$ for methane.
Since our system is idealized as a monodimensional reservoir, along the vertical direction, the chemical composition gradients are obtained 
by integrating the following equations:
\begin{equation}\label{equa_x1}
   \frac{\partial x_1}{\partial z} = - \frac{\scriptstyle D_1^P D_{22}^M - D_2^P D_{12}^M}{\scriptstyle D_{11}^M D_{22}^M - D_{21}^M D_{12}^M} \, 
                                                              \frac{\partial P}{\partial z} 
                                     - \frac{\scriptstyle D_1^T D_{22}^M - D_2^T D_{12}^M}{\scriptstyle D_{11}^M D_{22}^M - D_{21}^M D_{12}^M} \, 
                                                              \frac{\partial T}{\partial z} 
,\end{equation}
\begin{equation}\label{equa_x2}
   \frac{\partial x_2}{\partial z} = - \frac{\scriptstyle D_1^P D_{21}^M - D_2^P D_{11}^M}{\scriptstyle D_{12}^M D_{21}^M - D_{22}^M D_{11}^M} \, 
                                                            \frac{\partial P}{\partial z} 
                                     - \frac{\scriptstyle D_1^T D_{21}^M - D_2^T D_{11}^M}{\scriptstyle D_{12}^M D_{21}^M - D_{22}^M D_{11}^M} \, 
                                                              \frac{\partial T}{\partial z} 
.\end{equation}
  The abundance of species 3 (methane here) is implicitly computed according to the normalization condition of mole fractions. 
It is worth noting that the presence of methane is represented in the equation by physical quantities like molecular diffusion 
coefficient $\mathcal{D}_{i3}$ (see below) or the density of the liquid $\rho_{\rm liq,}$ which depends on the chemical composition.
All the detailed formulations of coefficients $D_{ij}^M$, $D_i^P,$ and $D_i^T$ are given in Appendix~\ref{ModelDiffTernary}.
%%%%% =====================
We emphasize that generalized diffusion coefficients, as they appear in Eqs~\ref{equa_x1} and \ref{equa_x2}, are noted with a ``$D$'' 
throughout the paper, while usual molecular diffusion coefficients are noted with a ``$\mathcal{D}$''.
%%%%% =====================
The temperature gradient sensitivity of molecular fluxes, represented by the $D_i^T$s, may be explicitly written
\begin{equation}
  D_i^T = a_{i3} \mathcal{D}_{i3} \bar{M} \frac{k_{T, \, i3}}{T}
.\end{equation}
The terms $a_{i3}$, $\mathcal{D}_{i3,}$ and $\bar{M}$ are defined in Appendix~\ref{ModelDiffTernary}, but their meaning is
not required for the discussion that follows. The thermal diffusion ratio $k_{T, \, i3}$ may be expressed as a function of the thermal diffusion
coefficient $\alpha_{T, \, i3}$. Unfortunately, the $\alpha_{T, \, i3}$s relevant for our purpose are not available in the literature, and their 
estimation is not a straightforward task.
This is why, in a first approach, we neglected the thermal diffusion contribution, as such the system is considered to be isothermal with a 
uniform temperature taken equal to the surface temperature $T_0= 90$~K. In such a situation, the derivative $\partial T/\partial z$
in Eqs~\ref{equa_x1} and \ref{equa_x2} vanishes, and the pressure derivative is provided by the hydrostatic equilibrium of the alkanofer:
\begin{equation}\label{hydroEqui}
  \frac{\partial P}{\partial z} = -\rho_{\rm eff} \, g_{\rm Tit}
,\end{equation}
where $\rho_{\rm eff}$ (kg~m$^{-3}$) is the effective density, which takes into account the matrix of water-ice, and $g_{\rm Tit}$ (m~s$^{-2}$) 
is Titan's gravity. If $\Pi$ is the average porosity of Titan's icy crust, then $\rho_{\rm eff}= \Pi \, \rho_{\rm liq} + (1-\Pi) \, \rho_{\rm ice}$ 
where $\rho_{\rm ice}$ is the density of ice I$_{\rm h}$ and $\rho_{\rm liq}$ the density of the cryogenic liquid mixture. 
We neglected the depth dependency of $\Pi$, which can be found in other works \citep{kossacki_lorenz_1996}, since porosity has no direct influence on molecular diffusion (see also the results reported at the end of this section).
Here, $\rho_{\rm liq}$ is computed in the frame of the PC-SAFT\footnote{Perturbed-Chain Statistical Associating Fluid Theory.} 
\citep{gross_sadowski_2001} equation of state (EoS), successfully used in several past studies dealing with the Titan context \citep{tan_etal_2013,luspay_kuti_etal_2015,cordier_etal_2016b,cordier_etal_2017a}. 
This equation of state is also employed to estimate the fugacity derivatives $\partial \mathrm{ln} f_i/\partial x_j$ and the partial molar volumes 
$\bar{V}_i$, which appear in the expressions of $D_{ij}^M$s and $\bar{V}_i$s (see Appendix~\ref{ModelDiffTernary}).
The water-ice density $\rho_{\rm ice}$ is evaluated according to a dedicated I$_{\rm h}$ ice equation of state \citep{feistel_wagner_2006}.\\

The fickean molecular diffusion coefficients' $\mathcal{D}_{ij}$s are needed by $D_{ij}^M$s' calculations. 
Given that we are dealing with liquids, we employed Wilke \& Chang's method \citep[hereafter WC55 method; see][]{wilke_chang_1955} 
combined with the \cite{batschinski_1913}, \cite{hildebrand_1971},
and \cite{vogel_weiss_1981} methods (nicknamed ``BHVW method'' in the following) to estimate the viscosity of liquids. 
The whole procedure is summarized in \cite{poling_2007} 
(see their Eqs.~9-11.6, p.~9.72). We stress that the WC55 method has a validity restricted to diluted solutions, whereas our studied mixtures 
are not necessary diluted. We discuss the influence of fickean
molecular diffusion coefficients on our results later in this paper. 
A few $\mathcal{D}_{ij}$ estimations at $P= 1.5$~bar and $T= 90$~K, obtained with WC55 approach lead to 
$\mathcal{D}_{\rm N_2-CH_4}= 1.6454 \times 10^{-9}$~m$^2$~s$^{-1}$ and
$\mathcal{D}_{\rm C_2H_6-CH_4}= 1.2439 \times 10^{-9}$~m$^2$~s$^{-1}$; these 
values around $10^{-9}$~m$^2$~s$^{-1}$ appear consistent with what we can expect for liquids.
%
%===================================================================================================================================
\begin{figure}[!t]
\begin{center}
\includegraphics[angle=0, width=8 cm]{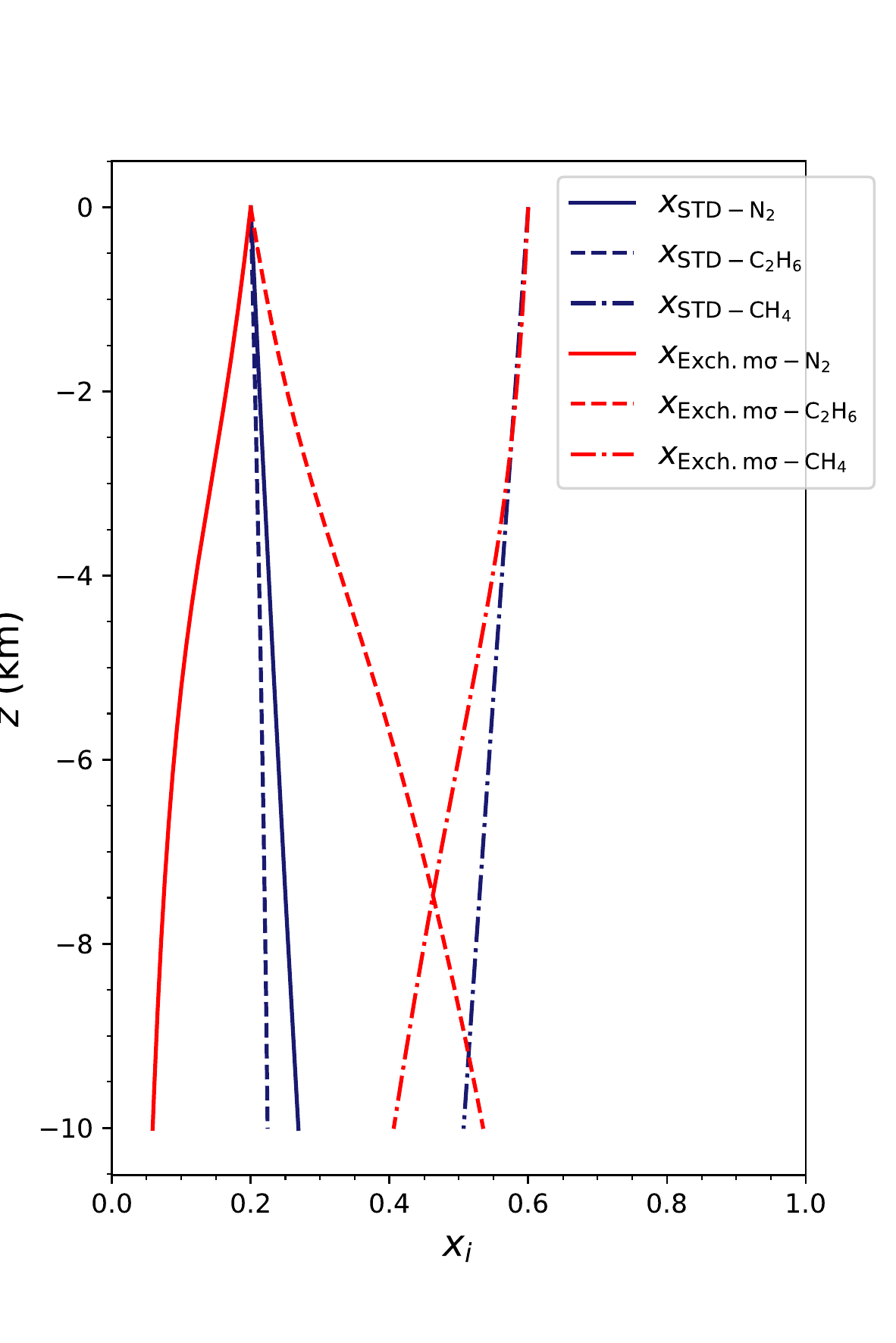}
\caption[]{\label{influ_mSigma}Influence of PC-SAFT parameters $m$ (number of segments) and $\sigma$ (segment diameter), representing the 
           size of the considered molecule. Our standard model, corresponding to Fig.~\ref{grad_isotherm_crust} panel (b), is in blue, while 
           red lines represent the model in which values of ($m$,~$\sigma$) for N$_2$ and C$_2$H$_6$ were exchanged. In order to avoid 
           any misinterpretation, we emphasize that the red lines have no physical meaning, they correspond only to a toy model dedicated to
           testing the sensivity of parameters.}
\end{center}
\end{figure}
%===================================================================================================================================
% 
  The set of differential equations governing the alkanofer abundance profile, Eqs~\ref{equa_x1}, \ref{equa_x2}, and 
\ref{hydroEqui}, is recognized as an initial value problem, which is also called the ``Cauchy problem''. Thus, at the surface, meaning at $z=0$, the pressure, 
temperature, and chemical composition have to be fixed and used as boundary conditions.

This numerical problem is solved using a standard Runge-Kutta algorithm \citep[e.g.,][]{nougier_1987}. 
For the surface pressure, we took Titan's ground pressure, 
measured in situ by the Huygens probe; it is close to $1.5$~bar \citep{fulchignoni_etal_2005} and should not vary significantly over the satellite 
surface. The temperature measured in a tropical region by Huygens is about $94$~K, but the temperature in polar regions, where lakes and seas are
located, should be a few degrees lower, that is around $90$~K \citep{jennings_etal_2016}. The chemical composition of seas remains relatively poorly 
constrained, this is the reason why we investigated several scenarios, keeping liquid methane as the solvent.\\
  In a first approach, with the aim of isolating the effect of gravity, and also because thermal diffusion coefficients are not well known, 
we chose to neglect the contribution of the geothermal gradient. This gradient is represented 
by the temperature derivative in Eqs.~\ref{equa_x1} and \ref{equa_x2}. As a consequence, the chemical composition gradients, represented by derivatives of compound 
mole fractions, are directly proportional to gravity. For instance,
\begin{equation}
   \frac{\partial x_1}{\partial z} \propto \rho_{\rm eff} g_{\rm Tit} 
,\end{equation}
then, for a low gravity object like Titan, where $g_{\rm Tit}= 1.352$~m~s$^{-2}$, we can expect a pretty smooth vertical stratification 
of chemical species.\\
%
% ----------------------------------------------------------------------------------
\begin{table}[!t]
%\begin{center}
\caption{\label{pcsaft_param}PC-SAFT parameters.}
%\vspace*{0.5cm}
\begin{tabular}{lllr}
\hline
Species     & $m$      & $\sigma$ (\AA)& $\epsilon/k_{\rm B}$ (K)   \\ 
N$_2$       & 1.2414   & 3.2992        &  89.2230                   \\
C$_2$H$_6$  & 1.6114   & 3.5245        & 190.9926                   \\
CH$_4$      & 1.0000   & 3.7039        & 150.0300                   \\
\hline
\end{tabular}
%\end{center}
\end{table}
% ----------------------------------------------------------------------------------
%
    The widely accepted thickness of Titan's crust should be in the range $80-100$~km 
\citep[see e.g.,][]{nimmo_bills_2010,tobie_etal_2012}, but the surface liquids are likely to be present only within the first few kilometers 
of the subsurface layers. To recall orders of magnitude, according to bathymetric maps, derived from Cassini Synthetic Aperture Radar observations 
\citep{hayes_2016}, the maximum depth of Titan's seas should be around $200$~m. In their subsurface liquid circulation model, \cite{horvath_etal_2016} 
assumed a scale height of $5$~km for the vertical hydraulic permeability variations law. On the Earth, fresh water aquifers seem to extend 
down to $\sim 2$~km below the ground, and salty water is also found in oil wells at depths below $\sim 6$~km \citep{baldwin_mcguinness_1976}. 
Then, we decided to fix the lower boundary of our model at $10$~km, a value that could reasonably represent the maximum
expected thickness of a Titan alkanofer. However, a well-defined value for this limit is not strictly required for our discussion.\\
%
%===================================================================================================================================
\begin{figure*}[!tp]
\begin{center}
\includegraphics[angle=0, width=18 cm]{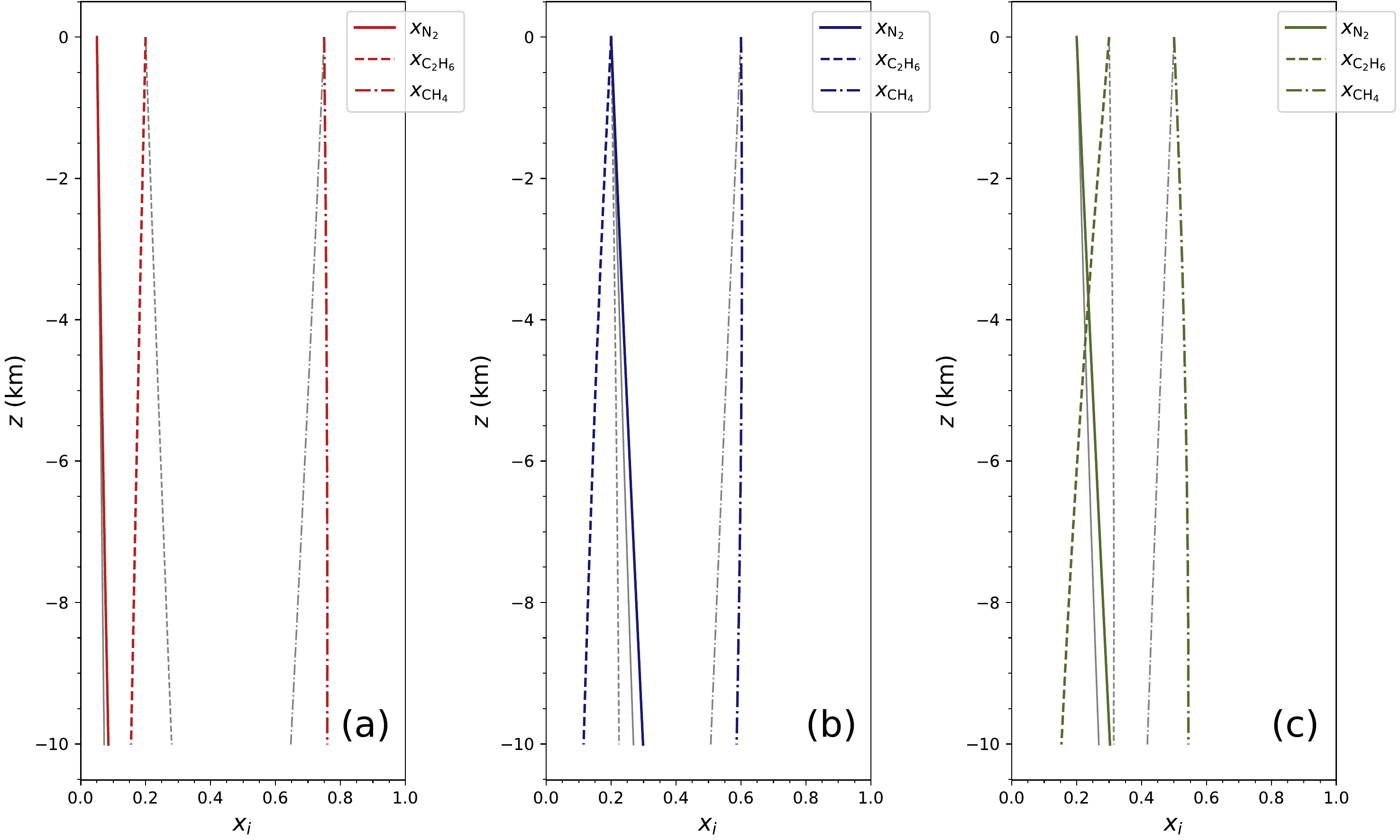}
\caption[]{\label{GeoTGrad_crust}Vertical variations of mole fractions (denoted $x_i$) of the main components (i.e., N$_2$,
           C$_2$H$_6$ and CH$_4$) of Titan's alkanofer liquid. This figure depicts a scenario where the icy crust undergoes a 
           geophysical thermal gradient derived from a model based on \cite{sohl_etal_2014}. While at the surface the temperature is 
           $T= 90$~K and the pressure has the ground value of $1.5$~bar, at a depth of $10$~km

           the pressure varies between $1.5$~bar at the surface ($z= 0$) and around $130$~bar 
           at the bottom of the simulated system (depth of $z= -10$~km). The three panels correspond to different surface compositions, 
           expressed in mole fractions: 
           (a) N$_2$: $0.05$, C$_2$H$_6$: $0.20$, CH$_4$: $0.75$;           (b) N$_2$: $0.20$, C$_2$H$_6$: $0.20$, CH$_4$: $0.60$;
           (c) N$_2$: $0.20$, C$_2$H$_6$: $0.30$, CH$_4$: $0.50$. For all computations, the porosity of the water ice matrix has been 
           fixed to 5\%, the diffusion theory formalism follows \cite{ghorayeb_firoozabadi_2000}, the water-ice EoS is provided by 
           \cite{feistel_wagner_2006}, while the thermodynamic properties of liquids are computed using PC-SAFT.
           For comparison, the results of Fig.~\ref{grad_isotherm_crust} are added in thin gray lines.}
\end{center}
\end{figure*}
%===================================================================================================================================
%
  The results, obtained within the theoretical framework described above, are plotted in Fig.~\ref{grad_isotherm_crust}. 
A first glance at Fig.~\ref{grad_isotherm_crust} reveals several striking features: 
(1) the progression of abundances with increasing depth is clearly linear; (2) the heaviest species tend to accumulate in the deepest layers as expected; 
(3) nitrogen (N$_2$: $28.0134$~g~mol$^{-1}$) seems to be more sensitive to gravitational effect than ethane (C$_2$H$_6$: $30.0690$~g~mol$^{-1}$), although 
ethane is slightly heavier than nitrogen. This latter aspect can only be explained by nonideal effects within the liquid solution. 
The PC-SAFT parameters recalled in Table~\ref{pcsaft_param} show that molecules are individually associated with a specific 
set of parameters: the number of segments $m$, the segment diameter $\sigma$ (\AA), and the segment energy $\epsilon$. 
   According to the paradigm of PC-SAFT theory, individual molecules are idealized by a collection of hard spheres (called ``segments''), 
and the parameters $m$ and $\sigma$ are the associated geometrical parameters. Of course, in realistic situations, individual molecules are not just a 
collection of identical hard spheres. Since the parameters $m$ and $\sigma$ are adjusted in order to fit experimental results, 
$m$ are real numbers, not necessarily integers. Moreover, interaction 
parameters, denoted $k_{ij}$, are introduced to account for inter-species interactions. Their values were taken from previous papers \citep{cordier_etal_2016b,tan_etal_2013}: 
$k_{\rm C_2H_6-N_2}= 0.07$,
$k_{\rm C_2H_6-CH_4}= 0.00$ and 
$k_{\rm N_2-CH_4}= 0.03$. We checked, by setting all these $k_{ij}$s to zero, that the interaction parameters have no influence on the obtained 
abundances profiles. Similarly, by increasing the value of $\epsilon$(N$_2$) to a value comparable to that of other species, we found the segment 
energy having no significant role. Finally, we exchanged the ($m$,~$\sigma$) tuple between N$_2$ and C$_2$H$_6$. For a given molecule,
the tuple ($m$,~$\sigma$) represents its size.
In Fig.~\ref{influ_mSigma}, we compare a standard model with a model where we switched these ($m$,~$\sigma$) values.
This operation has a huge effect on the resulting profiles, with ethane becoming the dominant compound in the deepest layers of the 
reservoir.
Therefore, the surprising behavior depicted in Fig.~\ref{grad_isotherm_crust}, leading to an alkanofer bottom more enriched in N$_2$ than 
in C$_2$H$_6$, is caused by nonideal effects due to the difference in size (see $m$ and $\sigma$ values in Table~\ref{pcsaft_param}) 
of these molecules. Although the nitrogen molecule is slightly lighter than the ethane one; the smaller size of nitrogen molecule overcomes the effect of gravity.\\

 To proceed to quantification, we introduced a vertical enrichment ratio for a given species $i:$
\begin{equation}\label{vertenrich}
   \Delta_i = \frac{x_{i, b} - x_{i, s}}{x_{i, s}}
,\end{equation}
with mole fractions $x_i$ subscripted with $b$ or $s$ corresponding to bottom or surface of the system. For the scenarios reported in 
Fig.~\ref{grad_isotherm_crust}, $\Delta_{\rm N_2}$ ranges between $34$\% and $45$\%, showing a clear enrichment. In contrast,
ethane undergoes 
a lower enrichment with $\Delta_{\rm C_2H_6}$ between $4.6$\% and $40$\%. For each scenario, the nitrogen enrichment is larger
than ethane one.\\

  We found that the surface pressure has only a negligible influence on the vertical composition profile. For instance, if we fix
this pressure to $3$~bar, a value that could be reached at Titan's sea bed \citep[see][]{cordier_etal_2017a}, the profile remains unchanged. 
Equivalently, we obtain a globally similar picture by varying the temperature between $85$ and $95$~K or by changing the assumed 
uniform porosity from $5$\% to $20$\%. The latter parameter has only an indirect influence via the effective density $\rho_{\rm eff}$ 
, which changes the value of the pressure through Eq.~\ref{hydroEqui}.

%%%%%%%%%%%%%%%%%%%%%%%%%%%%%%%%%%%%%%%%%%%%%%%%%%%%%%%%%%%%%%%%%%%%%%%%%%%%%%%%%%%%%%%%%%%%%%%%%%%%%%%%%%%%%%%%%%%%%%%%%%%%%%%%%%%%
%%%%%%%%%%%%%%%%%%%%%%%%%%%%%%%%%%%%%%%%%%%%%%%%%%%%%%%%%%%%%%%%%%%%%%%%%%%%%%%%%%%%%%%%%%%%%%%%%%%%%%%%%%%%%%%%%%%%%%%%%%%%%%%%%%%%
%
%\clearpage
\section{Contribution of thermal diffusion}
\label{nonisothermcrust}

  Due to the vertical geothermal gradient, fluids confined in terrestrial oil or gas reservoirs undergo thermodiffusion, which can have, 
over geological timescales, consequences on the segregation of chemical species comparable to that of pressure and gravity \citep{galliero_etal_2017}. 
In some cases, thermodiffusion shows even counterintuitive effects, leading heavy fluid mixtures floating on top of light fluids layers
\citep{ghorayeb_etal_2003}. Thus, it is important to tentatively estimate the possible amplitude of this effect in our context.

   Like many other moons, Titan has internal energy sources provided by radio elements and tidal effects \citep{tobie_etal_2006}.
This heat has to be dissipated and it is transported toward the surface through geological layers. According to numerical models 
\citep{sohl_etal_2014}, the flux at the surface should be $\sim 1$~mW~m$^{-2}$, to be compared with the average geothermal flux at the surface of
the Earth, which is more or less two orders of magnitude higher \citep{pollack_etal_1993}. The thermal structure of Titan's crust can be 
reconstructed (see the description of our model in Appendix~\ref{structcrust}), assuming an I$_{\rm h}$-ice viscosity \citep{sohl_etal_2014} high enough to 
impede solid-state convection \citep{nimmo_bills_2010}. The results show a linear evolution of the temperature within the layers 
of interest, corresponding to a gradient of $\sim 0.6$~K~km$^{-1}$, well below the terrestrial geothermal gradient that has a commonly 
accepted value around $30$~K~km$^{-1}$ \citep{lowrie_2010}. The existence of ice convection in Titan's outer crust is still debated 
\citep{nimmo_bills_2010,liu_etal_2016,kalousova_sotin_2020}. However, a recent work \citep{kalousova_sotin_2020} suggests the existence of 
some convection below $\sim 15$~km, allowing for the possible icy alkanofer to remain in a static state.\\
\indent  In Equations~\ref{equa_x1} and \ref{equa_x2}, the geothermal gradient terms involve two coefficients: $D^T_1$ and $D^T_2$,
which contain thermal diffusion coefficients (see Eq.~\ref{D1_T} and \ref{D2_T} in Appendix~\ref{ModelDiffTernary}). Here, we need two 
coefficients: for N$_2$ in CH$_4$ (denoted $\alpha_{\rm N_2-CH_4}$) and for C$_2$H$_6$ in CH$_4$ (denoted $\alpha_{\rm C_2H_6-CH_4}$), which 
are not available in the literature. In the general case, the thermal diffusion coefficient, $\alpha_{ij,T}$, of two species $i$ and $j$ 
is determined, in a complex manner, by the sizes and masses of molecules, the temperature and composition of the mixture, and the 
intermolecular interactions. This aspect is particularly relevant for nonideal fluids like those involved here. While 
performing accurate laboratory measurements of $\alpha_{ij,T}$s under microgravity conditions is difficult \citep{hu_etal_2014}, theoretical 
estimations are possible. Consistently with the first model described in the previous section, 
for species (1) and (2) $\alpha_{12, T}$ can be derived from PC-SAFT via Haase's formula \citep{haase_1969,pan_etal_2006}:
\begin{equation}\label{alphaHaase}
  \alpha^{\rm Haase}_{12, T} = \frac{RT}{x_1 \left(\frac{\partial\mu_1}{\partial x_1}\right)} \, 
     \left\{\alpha^0_{12,T} + \frac{M_1 \frac{\bar{h}_2^{res}}{RT} - M_2 \frac{\bar{h}_1^{res}}{RT}}{M_1 x_1 + M_2 x_2}\right\}
,\end{equation}
where the $M_i$s and $x_i$s are, respectively, the molecular weights and the mole fractions, and $\mu_1$ is the chemical potential of component ($1$).
The coefficient $\alpha^0_{12,T}$ represents the thermal diffusion coefficient for the corresponding ideal gas, $\bar{h}_1^{res}$ and $\bar{h}_2^{res}$
are the respective residual partial molar enthalpies of species (1) and (2). Even for most nonideal fluids $\alpha^0_{ij,T} \ll \alpha_{ij,T}$
\citep{pan_etal_2006}, the coefficients $\alpha^0_{ij,T}$ can be easily computed by applying the 
kinetic theory of gases \citep{chapman_cowling_1970}. On their side, the residual enthalpies are directly provided by PC-SAFT 
\citep[see Eq. A.46 of ][]{gross_sadowski_2001}:
%
%===================================================================================================================================
\begin{figure}[!t]
\begin{center}
\includegraphics[angle=0, width=8.5 cm]{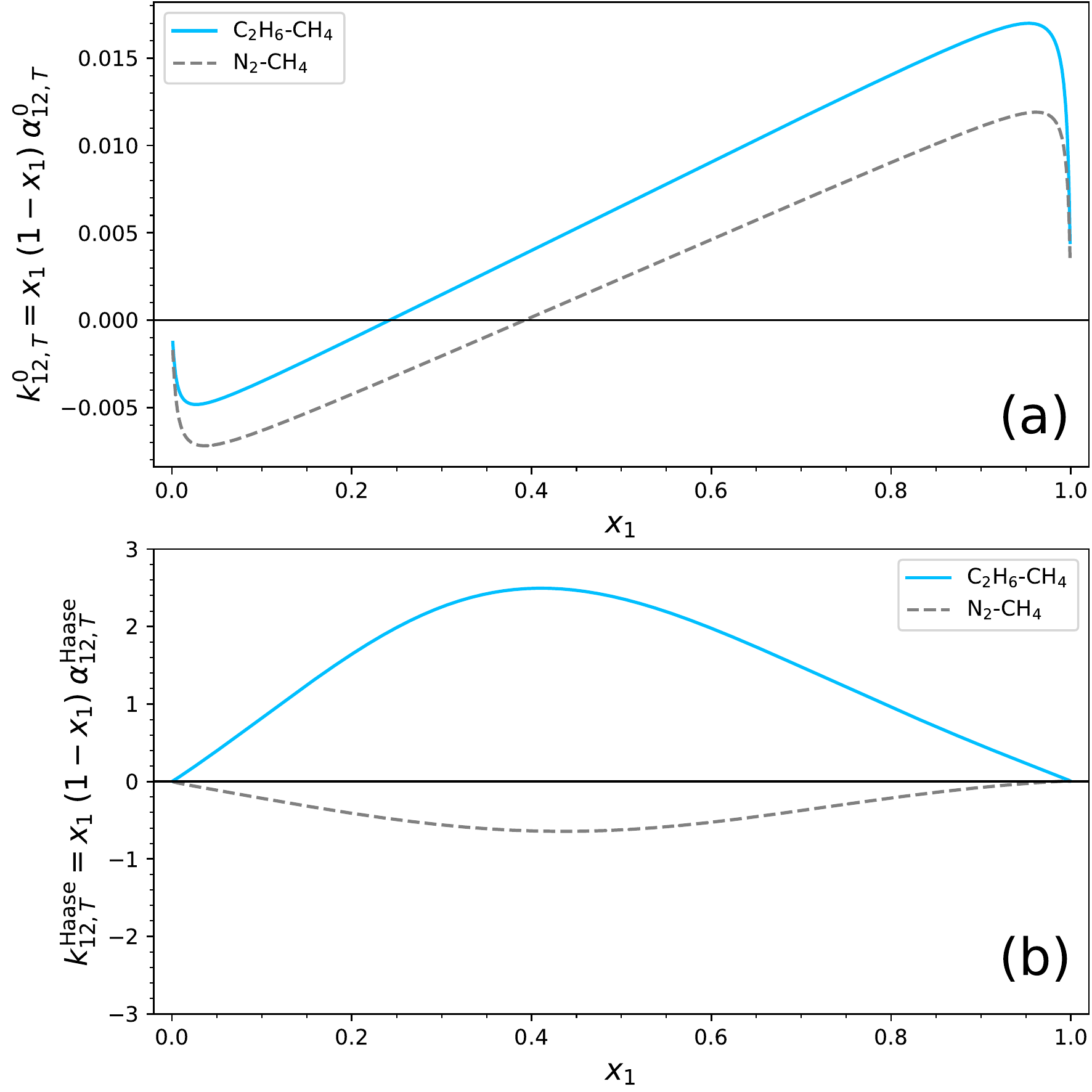}
\caption[]{\label{variaAlpha}(a) Variation of $k^0_{12,T}=x_1 (1-x_1) \, \alpha^0_{12, T}$ as a function of the mole 
           fraction $x_1$ of the first compound. The quantity $k^0_{12,T}$ is chosen, because it represents the sensitivity of the 
           system to a gradient of temperature. In this panel, all quantities are related to the ideal gas state (marked by 
           the ``$0$'' superscript).
           (b) Similar plot using the thermal diffusion coefficient provided by Haase's theory (see Eq.~\ref{alphaHaase}).
           In both panels, the solid line is associated with the C$_2$H$_6$-CH$_4$ system, while the dashed line is related to
           N$_2$-CH$_4$.
           All computations represented in this figure assume a pressure of $1.5$~bar and a temperature of $90$~K.}
\end{center}
\end{figure}
%===================================================================================================================================
% 
\begin{equation}
 \frac{h^{res}}{RT} = -T \left(\frac{\partial \tilde{a}^{res}}{\partial T}\right)_{\rho, x_i} + (Z-1)
,\end{equation}
where $\tilde{a}^{res}$ is the residual and reduced (i.e., molar) Helmholtz free energy, and $Z$ is the compressibility factor
\citep{gross_sadowski_2001}. The residual partial molar enthalpies are simply derived using
\begin{equation}
    \bar{h}^{res}_i = \left(\frac{\partial h^{res}}{\partial x_i}\right)_{P, T, x_{j \ne i}}
.\end{equation}
In our framework, the term ``residual'' relates to a difference between the physical quantities for the actual fluid 
and the corresponding ideal gas. In the case of the chemical potential derivative, we also have
\begin{equation}
  \frac{\partial \mu_i}{\partial x_i} = \frac{RT}{x_i} + R T \frac{\partial}{\partial x_i} \left(\frac{\mu_i^{res}}{k_{\rm B} T}\right)
,\end{equation}
with $k_{\rm B}$ being the Boltzmann constant. The dimensionless ratio $\mu^{res}/k_{\rm B} T$ is directly provided by PC-SAFT 
\citep[see Eq.~A.33 of][]{gross_sadowski_2001}.
Of course, for any physical quantity $Q$, its residual counterpart $Q^{res}$ approaches zero when 
the system approaches an ideal gas state. This way, for an ideal gas $\partial \mu_i^{\rm id. gas} /\partial x_i \sim RT/x_i,$ and consequently 
\begin{equation}
   \alpha^{\rm Haase, id. gas}_{12, T} \simeq \alpha^0_{12,T}
,\end{equation}
as expected. In Fig.~\ref{variaAlpha}, for illustration purposes, we report thermal diffusion factors $k_{ij, T}$ for the required pairs 
of species: N$_2$-CH$_4$ and C$_2$H$_6$-CH$_4$. Except for $x_1$ near zero or one, the sensitivity of nonideal 
systems (panel b) to temperature, represented by $k_{ij, T}$, is larger than the sensitivity of ideal systems (panel a). Moreover, according to
these estimations, ethane seems to be more sensitive than nitrogen: 
$|k_{{\rm C_2H_6-CH_4}, T}^{\rm Haase}| \ge |k_{{\rm N_2-CH_4}, T}^{\rm Haase}|$. If we consider only thermodiffusion in a binary system, 
the flux $j_1$ of the species ($1$) along the vertical $z$-axis is given by \citep[see Eq. 1 in][]{cordier_etal_2019}
\begin{equation}
 j_1 = -\frac{k_{12, T}}{T} \frac{\partial T}{\partial z}
.\end{equation}
In the case of our icy crust model, we found  $|\partial T/\partial z| \sim 0.6$~K~km$^{-1}$; since 
$k_{{\rm C_2H_6-CH_4}, T}^{\rm Haase} \ge 0$
and $k_{{\rm N_2-CH_4}, T}^{\rm Haase} \le 0$, we have $j_{\rm C_2H_6} \ge 0$ and  $j_{\rm N_2} \le 0$. As a consequence, the introduction 
of thermodiffusion should reinforce the abundance of nitrogen in the deepest parts of Titan's alkanofer. Of course, this approach is 
very simplified, if not naive, as it ignores the presence of the third species present in a ternary mixture, which we take into account in our full model.\\
In Fig.~\ref{GeoTGrad_crust}, the profiles of molecular abundances obtained for the models
including the geothermal gradient (colored lines) are compared to 
previous models corresponding to an isothermal icy matrix (gray lines). Clearly, the simple prediction made above seems to explain the results 
depicted in Figs.~\ref{GeoTGrad_crust}.b and \ref{GeoTGrad_crust}.c, corresponding to scenarios where nitrogen has a relatively high surface-mole fraction
(i.e., $x_{{\rm N_2},s}= 0.20$).
The first scenario, corresponding to panels (a) in the same figure, show a different behavior due to the low value of $x_{{\rm N_2},s}$ ($0.05$).
In this case, ethane can undergo a substantial enrichment in the lowest layers of the 
alkanofer. This result may be regarded as rather obvious, the competition between nitrogen and ethane being minimized. Again, 
if the amount of dissolved nitrogen is not too small, the bottom of the alkanofer is enriched in nitrogen.\\

\begin{figure}[!t]
\begin{center}
\includegraphics[angle=0, width=7 cm]{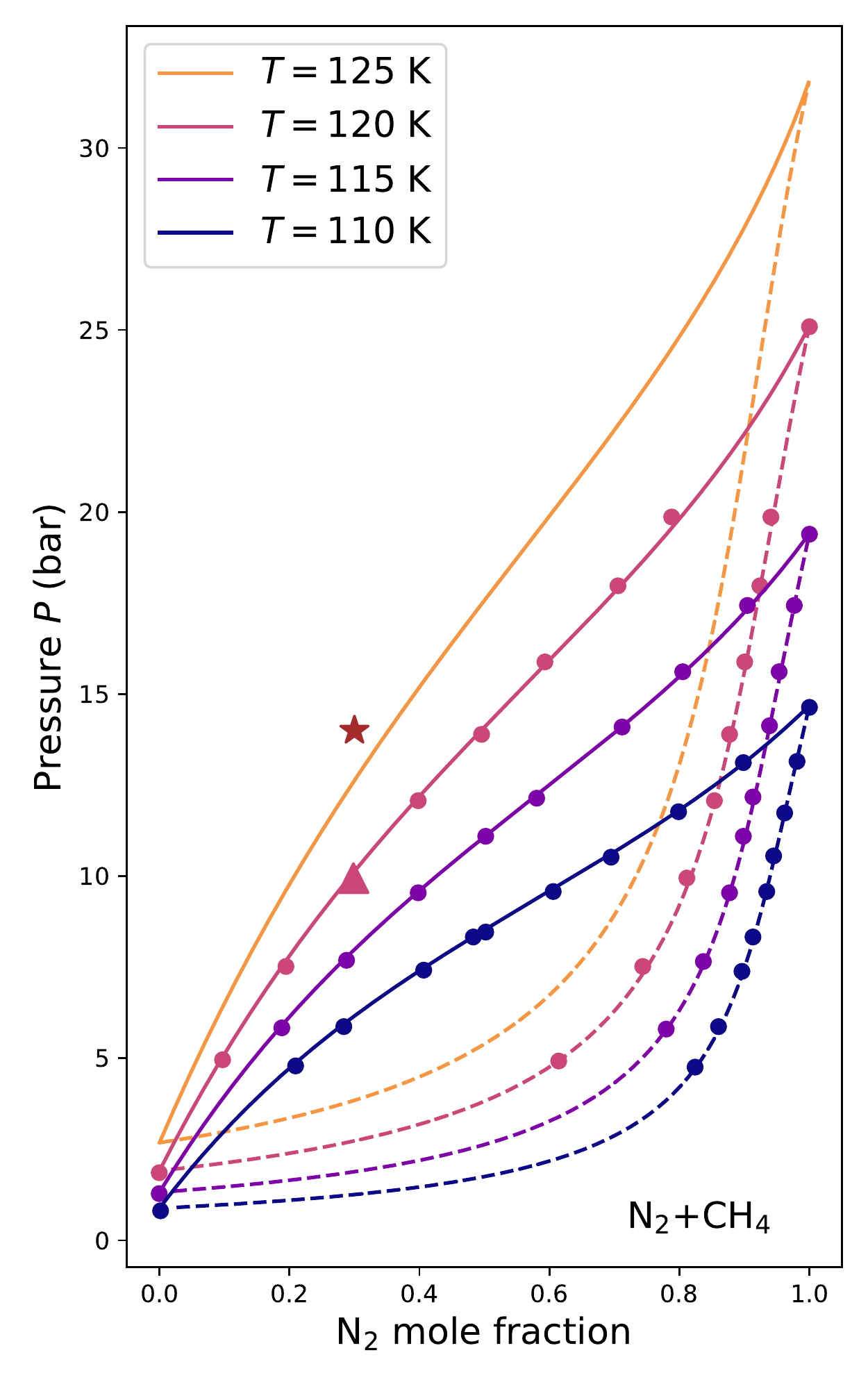}
\caption[]{\label{BinDiagN2CH4}Phase diagrams of the N$_2$+CH$_4$ system at $110$~K, $115$~K, $120$~K, and $125$~K.
           Circles represent laboratory measurements from \cite{parrish_hiza_1974}, curves are computed according to the PC-SAFT EoS
           (no experimental data are available at 125~K).
           Solid lines are the liquidus, while dashed lines represent the vaporus. The star represents the case where 
           $x_{\rm N_2}= 0.30$ and $P= 14$~bar, which is to be compared with the point marked by a triangle (see text).
           This figure is similar to Fig.~3b of \cite{tan_etal_2013}.}
\end{center}
\end{figure}

   As part of their work, \cite{kalousova_sotin_2020} built a 2D model of Titan's icy crust. These authors assumed the existence 
of a layer of clathrates at the surface of the crust. Since clathrates have a lower thermal conductivity than regular water ice, the 
geothermal gradients they found are much higher than those considered up to this point. Depending on the value of the 
width $h_c$ of the clathrated layer, \cite{kalousova_sotin_2020} obtained $|\partial T/\partial z|$ ranging between $2.8 $~K~km$^{-1}$ 
and $23.8 $~K~km$^{-1}$. Included in our model, such high gradients reinforce the effect already seen in Fig~\ref{GeoTGrad_crust} and 
discussed above, that is an enrichment in nitrogen and depletion in ethane. As an example, starting at the surface with a composition 
of 20\% of N$_2$, 20\% of C$_2$H$_6,$ and 60\% of CH$_4$, and assuming $\partial T/\partial z \sim -23.8$~K~km$^{-1}$ 
\citep[corresponding to $h_c = 5$~km in][]{kalousova_sotin_2020}, we obtained $x_{\rm N_2} \sim 0.30$ and $x_{\rm C_2H_6} \sim 0.0$ at the depth
of $1$~km. Below this depth, our model is no longer valid since it is dedicated to ternary mixtures. This example raises the question 
of the partial vaporization of the mixture. In Fig.~\ref{BinDiagN2CH4}, we plot a phase diagram of the binary system N$_2$+CH$_4$
for temperatures ranging from $110$~K to $125$~K. In the crust, at a depth of $1000$~m, the pressure should be around $14$~bar, and 
the temperature, according to Kalousova~\&~Sotin's gradient, should be between $114$~K and $120$~K. As we can see in Fig.~\ref{BinDiagN2CH4}, the 
point $(x_{\rm N_2} \sim 0.30, P= 14 \, {\rm bar})$, indicated by a star, is not very far from the liquidus at $120$~K (triangle in Fig.~\ref{BinDiagN2CH4}).\\
  Even if our estimations do not favor the appearance of nitrogen enriched bubbles in the possible Titan alkanofer, the question 
has to be kept in mind for future studies. Indeed, a subsurface degassing process could be the origin of geological activities. We can 
imagine the formation of chemical composition gradients, driven over geological timescales by diffusion, feeding the formation of some 
pockets of nitrogen which could, at some point, be released abruptly bringing other materials to planetary surface. This 
kind of mechanism may explain morphological evidence for volcanic activity in Titan's north polar regions \citep{wood_radebaugh_2020}.
%
%===================================================================================================================================
%===================================================================================================================================
%

%%%%%%%%%%%%%%%%%%%%%%%%%%%%%%%%%%%%%%%%%%%%%%%%%%%%%%%%%%%%%%%%%%%%%%%%%%%%%%%%%%%%%%%%%%%%%%%%%%%%%%%%%%%%%%%%%%%%%%%%%%%%%%%%%%%%
%%%%%%%%%%%%%%%%%%%%%%%%%%%%%%%%%%%%%%%%%%%%%%%%%%%%%%%%%%%%%%%%%%%%%%%%%%%%%%%%%%%%%%%%%%%%%%%%%%%%%%%%%%%%%%%%%%%%%%%%%%%%%%%%%%%%
%
%\clearpage
\section{Role of molecular diffusion coefficients}

  According to the formalism developed by \cite{ghorayeb_firoozabadi_2000}, 
which we adopted in the present work, the molecular ``fickean''
diffusion is represented by molecular diffusion coefficients, $\mathcal{D}_{ij}$s, which are key quantities. In the case of our ternary mixture, 
only two of them are explicitly included in the set of equations employed here: $\mathcal{D}_{13}$ and $\mathcal{D}_{23}$, 
corresponding, respectively, to the diffusion of 
N$_2$ and C$_2$H$_6$ in liquid methane. In models described in Sect.~\ref{isothermcrust} and Sect.~\ref{nonisothermcrust}, the values of 
these coefficients are estimated using the approximation proposed by \cite{wilke_chang_1955}, in this section we discuss the possible 
influence of these molecular diffusion coefficients.\\

   A quick look at the expressions of the Onsager phenomenological coefficients (see Eqs.~\ref{L11}, \ref{L22}, and \ref{L12}) leads 
to $L_{11} \propto \mathcal{D}_{13}$, $L_{22} \propto \mathcal{D}_{23}$ and $L_{12}=L_{21} \propto \mathcal{D}_{13}$, respectively. 
As a consequence, the ratio $L_{12}/L_{11}$ is independent of all molecular diffusion coefficients, whereas $L_{21}/L_{22}$ depends 
on $\mathcal{D}_{13}/\mathcal{D}_{23}$ since $L_{21}/L_{22} \propto \mathcal{D}_{13}/\mathcal{D}_{23}$. 
Therefore, the terms $D_{21}^M$ and $D_{22}^M$ are functions of the ratio $\mathcal{D}_{13}/\mathcal{D}_{23}$ and of $\mathcal{D}_{23}$, respectively.
This way, the equations that structure the diffusive properties of our model, namely Eqs.~\ref{equa_x1} and \ref{equa_x2}, have their 
barodiffusion and thermodiffusion terms as functions of $\mathcal{D}_{13}/\mathcal{D}_{23}$. Nonetheless, these dependencies are different since 
$D_i^P$ and $D_i^T$ have different behaviors regarding molecular diffusion coefficients (see Eqs.~\ref{D1_P}, \ref{D2_P}, \ref{D1_T}, and 
\ref{D2_T}). These dependencies have a quantitatively significant influence on derived chemical vertical profiles only if 
the values of other thermodynamical quantities do not inhibit this possible influence. This is why we performed sensitivity 
tests, multiplying $\mathcal{D}_{\rm N_2-CH_4}$ and $\mathcal{D}_{\rm C_2H_6-CH_4}$ by an arbitrary large factor. Setting this factor 
to $10^3,$ we found no influence on results concerning the ``isothermal crust'', whereas a relatively small influence
arises when we take into account the geophysical thermal gradient. The variation of the vertical enrichment ratio (see Eq.~\ref{vertenrich}) is about a
few tenths of a percent for nitrogen. The abundance profile of ethane seems to be more sensitive, with $\Delta_{\rm C_2H_6}$ going from 
$-42$\% (for our ``standard'' model) to $-39$\% (case $\mathcal{D}_{\rm C_2H_6-CH_4} \times 10^3$) or $-28$\% 
(case $\mathcal{D}_{\rm N_2-CH_4} \times 10^3$).\\

  In order to investigate the actual influence of these molecular diffusion 
coefficients more precisely, we also evaluated the diffusion coefficients of N$_2$ and C$_2$H$_6$ in CH$_4$ by molecular dynamics (MD) simulations.
A similar technique has already been employed, in the context of Titan, by \cite{kumar_chevrier_2019}.
Here, we have employed the well-known open-source package
GROMACS\footnote{GROningen MAchine for Chemical Simulations}$^{,}$\footnote{\url{http://www.gromacs.org}} 2018 version
\citep{abraham_etal_2015}. Nitrogen molecules have been described by the 
TraPPE\footnote{Transferable potentials for phase equilibria} force field \citep{potoff_siepmann_2001}. 
Alkanes (i.e., methane and ethane) were modeled 
by adopting the TraPPE-UA\footnote{Transferable potentials for phase equilibria - united atom} coarse-grained representation 
where pseudoatoms are introduced for methane and ethane methyl groups.
\citep{martin_siepmann_1998,mansi_etal_2017}.
For all computations, the intermolecular interaction cutoff
has been fixed to $1.4$~nm, since it appears to be a well -suited value \citep{martin_siepmann_1998,shah_etal_2017}.
We split our MD calculations into three steps: (1) a 1-ns equilibration step in the $NVT$ ensemble; (2) a 19-ns equilibration step 
in the $NpT$ ensemble, where the temperature and pressure are fixed to the expected target values; (3) a 10 ns-accumulation step in the $NpT$ 
ensemble.
In our work, standard simulations were performed with a total number of molecules fixed to $5000$ with $4000$ CH$_4$ (80\%) and $1000$ 
N$_2$ or C$_2$H$_6$ (20\%), the infinite spatial extent of the liquid being reproduced by imposing periodic boundary conditions to the 
simulation box in the three space directions.
\begin{table}[h]
%\begin{center}
\caption{\label{DijMD}Diffusion coefficients derived from our MD simulations based on TraPPE-UA force fields.\label{tab-Dbin}}
%\vspace*{0.5cm}
\begin{tabular}{llcccc}
      Species 1   &       Species 2 & $T$       & $P$       & \multicolumn{2}{c}{$D_{12}$} \\
                  &                 &           &           &        MD                    & WC \\
                  &                 & (K)       & (bar)     & \multicolumn{2}{c}{(10$^{-9}$ m$^2$ s$^{-1}$)} \\ 
\hline \\
N$_2$      (20\%) & CH$_4$  (80\%)  & 90        & 1.5       & $2.58\pm 0.10$               & $1.65$  \\
                  &                 & 95        & 120       & $2.81\pm 0.04$               &         \\ \\
C$_2$H$_6$ (20\%) & CH$_4$  (80\%)  & 90        & 1.5       & $1.06\pm 0.05$               & $1.24$  \\
                  &                 & 95        & 120       & $1.14\pm 0.02$               &         \\ 
\end{tabular}
%\end{center}
\end{table}
  In Table~\ref{DijMD}, we compare molecular diffusion coefficients derived from MD simulations, with estimations 
performed with Wilke \& Chang's (WC) method (see Sect.~\ref{isothermcrust}). The differences between the coefficients obtained with the MD and WC methods 
remain below a factor of two, which is clearly not enough to significantly alter the alkanofer chemical 
vertical profiles.
We can safely conclude that molecular diffusion coefficients have no influence on these profiles.

%%%%%%%%%%%%%%%%%%%%%%%%%%%%%%%%%%%%%%%%%%%%%%%%%%%%%%%%%%%%%%%%%%%%%%%%%%%%%%%%%%%%%%%%%%%%%%%%%%%%%%%%%%%%%%%%%%%%%%%%%%%%%%%%%%%%
%%%%%%%%%%%%%%%%%%%%%%%%%%%%%%%%%%%%%%%%%%%%%%%%%%%%%%%%%%%%%%%%%%%%%%%%%%%%%%%%%%%%%%%%%%%%%%%%%%%%%%%%%%%%%%%%%%%%%%%%%%%%%%%%%%%%
%
%\clearpage
\section{Discussion and conclusion}

%...................................................................................................................................
\subsection{Liquid circulation between subsurface reservoirs}

  In this work, the most important assumption is the absence of macroscopic flows between underground 
reservoirs, and inside the considered reservoir. More precisely, the timescales associated with convection and flows through porous 
media have to be much larger than timescales related to diffusive processes, like those discussed in previous sections. The order of 
magnitude of the ``diffusion timescale'', here denoted $\tau_{\rm diff}$, may be easily estimated using a relation similar to
\begin{equation}
   \tau_{\rm diff} \sim \frac{L^2}{\mathcal{D}}
,\end{equation}
with $L$ (m) being a characteristic length of the studied system, and $\mathcal{D}$ (m$^2$~s$^{-1}$) a diffusion coefficient. According to values 
reported in Table~\ref{DijMD}, the diffusion coefficient is on the order of $\sim 10^{-9}$~m$^2$~s$^{-1}$. For 
an alkanofer of depth $L \sim 100$~m, these values lead to a timescale of $\tau_{\rm diff} \sim 0.3$~mega-year, and 
$\tau_{\rm diff} \sim 30$~mega-years for a $1$~km deep reservoir. These timescales are comparable to those computed for terrestrial
hydrocarbon reservoirs \citep{obidi_2014}.\\

   According to Darcy's law \citep{darcy_1856}, the volume of liquid $\Delta V$ (m$^3$) exchanged between two reservoirs, 
with free surfaces altitudes difference of $h$, along a subsurface ``channel'' through a porous medium of length $L$ and cross section 
$S$, is given by
\begin{equation}\label{DeltaV}
  \Delta V = K_{\rm p, sc} \frac{\rho g S}{\eta L} \, h \, \Delta t
,\end{equation}
where $K_{\rm p, sc}$ is the permeability (m$^2$) of the porous medium that connects two reservoirs or a reservoir and a lake, and
the subscript ``sc'' stands for ``subsurface channel''.
In Eq.~\ref{DeltaV}, $\rho$ is the density of the liquid (kg~m$^{-3}$), $g$ is the 
gravity (m~s$^{-2}$), $\eta$ is the dynamic viscosity (Pa~s), $\Delta t $ is the time required to obtain free surfaces at the 
same equipotential. In other words, the transport of the volume $\Delta V$ of liquid has the duration $\Delta t $. If we consider 
two reservoirs with identical permeability and geometries, one in a terrestrial context, the second on Titan, the relationship between 
Earth and Titan timescales is given by
\begin{equation}
   \Delta t_{\rm Titan} = \frac{\eta_{\rm Titan}}{\eta_{\rm Earth}} \times 
                          \frac{\rho_{\rm Earth} \, g_{\rm Earth}}{\rho_{\rm Titan} \, g_{\rm Titan}} \times
                          \Delta t_{\rm Earth}
.\end{equation}
For the Earth, the working liquid is water, and in the case of Titan we considered a typical mixture 
N$_2$+CH$_4$+C$_2$H$_6$. We found
\begin{equation}
   \Delta t_{\rm Titan}  \sim 3 \, \Delta t_{\rm Earth}
,\end{equation}
a result that is in full accordance with an estimation made by \cite{horvath_etal_2016}. This longer timescale clearly favors the formation 
of liquid pockets, isolated from a large-scale subsurface environment. However, it is difficult to give a realistic estimate of 
horizontal seepage between two alkanofers, or between a lake and an alkanofer. In their work, \cite{hayes_etal_2008} computed an 
interaction timescale $\tau_{1/2}$ (see their Eq.~2) that depends on geometrical factors and on the terrains permeability if we put
aside the role of surface evaporation. Nonetheless, we can reformulate Eq.~\ref{DeltaV} as
\begin{equation}\label{DeltaT}
  \Delta t = \frac{1}{K_{\rm p, sc}} \, \frac{\Delta V}{h} \, \frac{L}{S} \, \frac{\eta}{\rho g}
,\end{equation}
where the ratio $\eta/\rho g$ represents the properties of the fluid, in the context of Titan ($g$), for this term we found 
roughly $10^{-8}$~m~s, since the dynamic viscosity $\eta$ may be computed by using the BHVW method, already mentioned 
in Sect.~\ref{isothermcrust}. The factor $L/S$ represents the geometry of the ``subsurface channel'', through which the liquid is 
supposed to circulate. If we take plausible values, for instance $L \sim 1$~km and $S \sim 1$~km$^2$, then 
$L/S \sim 10^{-3}$~m$^{-1}$. Moreover, we can consider that the composition of the targeted alkanofer should be significantly 
modified if $\Delta V$ is large enough; one can take, for instance, $\Delta V \sim h^3$, letting the ratio $\Delta V/h$ 
be comparable to $h^2$. On Titan, the topography is rather flat, $h \sim 1$~km is a large value. Keeping this value, we obtain 
$\Delta V/h \sim h^2 \sim 10^6$~m$^2,$ which leads to
\begin{equation}
  \Delta t = \frac{1}{K_{\rm p, sc}} \, 10^{-5}
.\end{equation}
Thus, the effect of diffusion is not erased by inter-reservoir seepage if, roughly,  $K_{\rm p, sc} \lesssim 10^{-5}/\Delta t$.
For a diffusion timescale, for example $\Delta t \sim 1$~mega-year (i.e., $3 \times 10^{13}$~s), 
we have $K_{\rm p, sc} \lesssim 3 \times 10^{-19}$~m$^2$. This result corresponds to impervious geological layers \citep[see Table~5.5.1 in][]{bear_1972}, which 
should surround the alkanofer, leaving the liquid trapped in a pocket for geological timescales. This possibility is compatible with topographic data analysis
by \cite{hayes_etal_2017}, who observed lakes or dry lake beds at elevations that require either a localized aquifer or a low permeability.
Nevertheless, geometrical parameters, presented in Eq.~\ref{DeltaT}, are very uncertain and may conspire to yield a $K_{\rm p, sc}$ value orders of magnitude 
higher or lower than what we find.

%...................................................................................................................................
\subsection{Convection in trapped liquids}

   When a fluid trapped in a porous matrix undergoes a temperature gradient, 
convection appears and the thermal convection flux overcomes the viscous forces; the ratio between these quantities 
is measured by the Rayleigh number ($\mathcal{R}{\rm a}$) \citep{bories_cambarnous_1973}:
\begin{equation}\label{RaNumb}
  \mathcal{R}{\rm a} = \frac{\alpha_{\rm V} \, K_{\rm p, m} \, g_{\rm Tit} \, (\rho C_{\rm p})_{\rm fluid} \, 
                        H^2}{\nu \lambda_{L}} \, \frac{\Delta T}{H}
,\end{equation}
with $\alpha_{\rm V}$ being the volumetric thermal expansion coefficient (K$^{-1}$), $K_{\rm p, m}$ the permeability of the reservoir 
matrix (m$^2$), $g$ the acceleration due to gravity (m~s$^{-2}$), $(\rho C_{\rm p})_{\rm fluid}$ the product of the fluid density (kg~m$^{-3}$) and the heat 
capacity (J~K$^{-1}$~kg$^{-1}$), $H$ the thickness of the reservoir (m), $\nu$ the kinematic viscosity of the fluid 
(Pa~s~kg$^{-1}$~m$^{3}$, i.e., m$^2$~s$^{-1}$), and $\lambda_{L}$ the 
thermal conductivity (W~m$^{-1}$~K$^{-1}$) of the medium. It is accepted that convection occurs when $\mathcal{R}{\rm a} \gtrsim 4\pi^2$
\citep{nield_bejan_2017}. The volumetric coefficient of thermal expansion $\alpha_{\rm V}$ is given by 
\begin{equation}
   \alpha_{\rm V} =  \frac{1}{V_{\rm m}} \, \left(\frac{\partial V_{\rm m}}{\partial T}\right)_{P} 
                  = -\frac{1}{\rho} \, \left(\frac{\partial \rho}{\partial T}\right)_{P} 
,\end{equation}
with $V_{\rm m}$ the molar volume of the fluid, and $\rho$ the density. 
For a given composition, pressure, temperature, heat capacity $C_{\rm P,}$ and density can be provided by PC-SAFT. 
The thermal conductivity $\lambda_{L}$ can be evaluated according 
to \cite{latini_etal_1996} and \cite{poling_2007} for saturated hydrocarbons, and following \cite{powers_etal_1954} for liquid nitrogen.
Finally, the kinematic viscosity $\nu$ is easily derived from dynamic viscosity $\eta = \rho \nu$, which is estimated according 
to the BHVW method mentioned in Sect.~\ref{isothermcrust}. 
%%%%%%%%%%%%%%%%%%%%%%%%%%%%%%%%%%%%%%%%%%%%==============================================%%%%%%%%%%%%%%%%%%%%%%%%%%%%%%%%%%%%%%%%%%%%%%%%%%
  As a robustness check for the validity of estimations performed accordingly to the above-mentioned methods, we also derived values of $C_{\rm P}$,  $\alpha_{\rm V,}$ and  $\eta$ from 
MD simulations on the ternary mixture CH$_4$(60\%)-C$_2$H$_6$(20\%)-N$_2$(20\%).
The outputs of these compulations, made in the frame of the isothermal-isobaric ($NpT$) ensemble, have been processed
according to relevant fluctuation formulas, leading to results similar to those obtained with PC-SAFT and other cited methods. The latter have 
been adopted for routine calculations since they are much more computationally flexible and efficient.
%%%%%%%%%%%%%%%%%%%%%%%%%%%%%%%%%%%%%%%%%%%%==============================================%%%%%%%%%%%%%%%%%%%%%%%%%%%%%%%%%%%%%%%%%%%%%%%%%%
For an alkanofer $1$~km deep and containing a typical mixture made up of 
$20$\% N$_2$, $60$\% CH$_4,$ and $20$\% C$_2$H$_6,$ under a thermal gradient of $0.6$~K~km$^{-1}$, we found $\mathcal{R}{\rm a} \sim 3.5$
(i.e., $\mathcal{R}{\rm a}/4\pi^2 < 1)$ \citep{nield_bejan_2017} for a permeability of $K_{\rm p, m}= 10^{-12}$~m$^2$, 
equivalent to that of very fine sand on Earth
\citep[see Table~5.5.1 in][]{bear_1972}. Within the terrestrial context, the empirical relationship between grain diameter $d$ (in micron) and 
$K_{\rm p, m}$ are available in the literature \citep{bear_1972}, leading to $d \simeq 40$~$\mu m$. Of course, this kind of estimation 
must be regarded with caution, since empirical laws established for the Earth case may no longer be valid for Titan, even if it provides 
a plausible value.
%
%===================================================================================================================================
\begin{figure}[!t]
\begin{center}
\includegraphics[angle=0, width=8.5 cm]{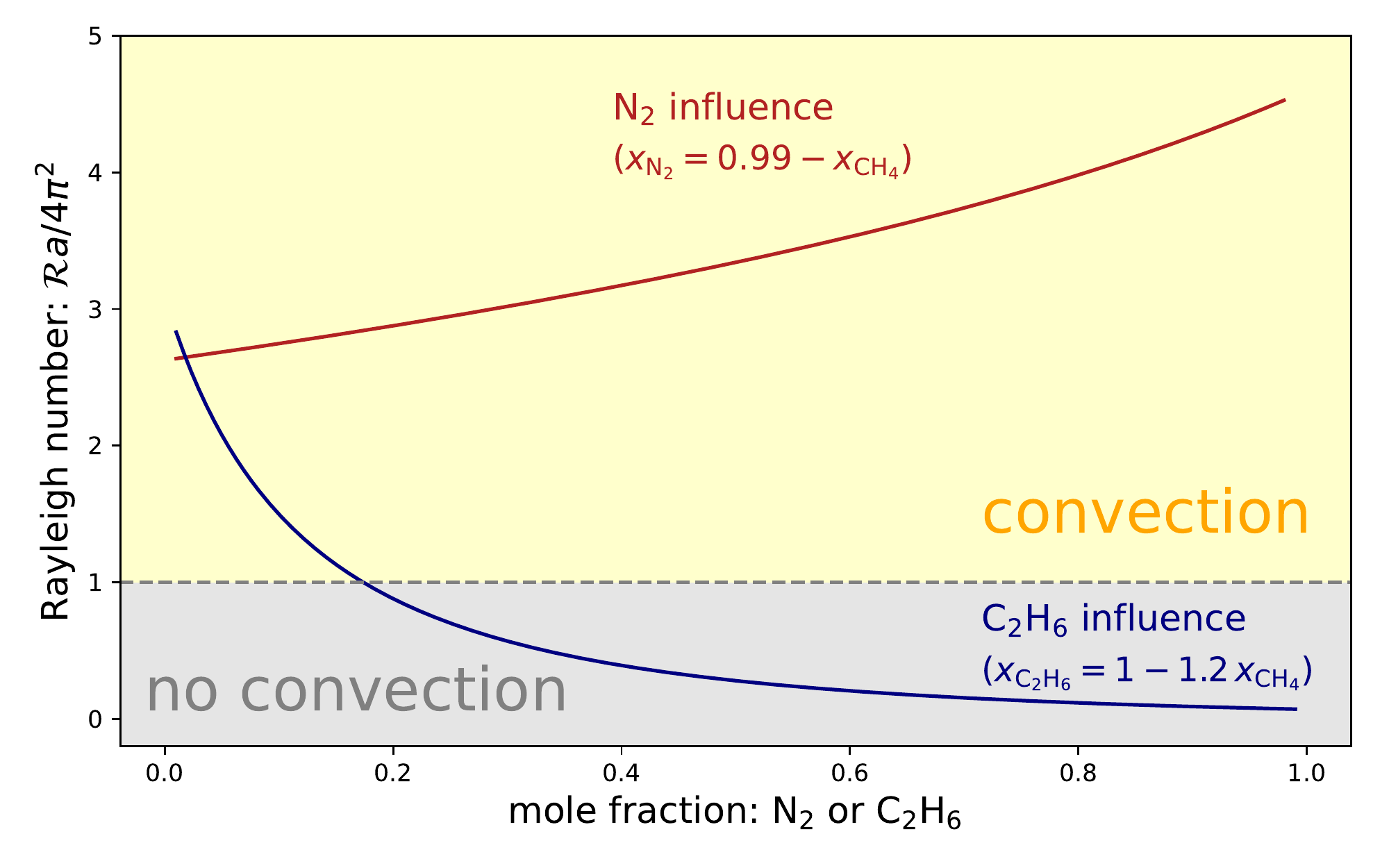}
\caption[]{\label{influChemCompoRa}Influence of the chemical composition of the fluid on its Rayleigh number. These computations were made 
           with a fixed permeability $K_{\rm p, m}= 10^{-12}$~m$^2$, $P= 20$~bar, $T= 90$~K, $H = 1$~km, and a thermal gradient 
           of $0.6$~K~km$^{-1}$.
           The influence of ethane content (blue curve) has been explored by forcing the mole fraction of nitrogen to follow that of 
           methane: $x_{\rm N_2} = 0.20 \, x_{\rm CH_4}$. The role of nitrogen abundance (red curve) has been studied by keeping ethane 
           mole fraction constant: $x_{\rm C_2H_6}= 0.01$.}
\end{center}
\end{figure}
%===================================================================================================================================
%
   Equation~\ref{RaNumb} clearly shows that low permeability $K_{\rm p, m}$, small depths $H,$ and gentle thermal gradients 
$\Delta T/H$ are favorable factors for the absence of convective transport in a Titan alkanofer. The role of chemical composition 
does not appear explicitly, but physical quantities such as $\alpha_{V}$, $\rho$, $C_{\rm P}$, $\nu,$ and $\lambda_{L}$ depend on the composition. 
In Fig.~\ref{influChemCompoRa}, for typical values of these physical parameters, we varied the mole fractions of nitrogen and 
ethane independently. A high abundance of ethane would damp convective fluxes, while large 
nitrogen abundances would have the opposite effect.\\
 Finally, the low permeability requirement has implications concerning the formation of a ``diffusive alkanofer''. Liquids of course have to fill
the reservoir either following a ``diffusive regime'' if the permeability is initially low, or following a ``hydrodynamic regime'' if the 
permeability is relatively large during the early ages of the reservoir. In the latter case, the permeability has to decrease with time
due to geophysical processes like tectonic activity or sedimentary deposits; this way, the permability could reach values low enough to damp convection.

   In this work, the chemical composition of the solid matrix 
has been hypothesized to be I$_{\rm h}$-water ice. Of course, even if this kind of scenario is very plausible and supported by our knowledge of 
Titan's inner structure, the actual Titan reservoirs could be somewhat different in nature. Carbon -bearing species seem to be 
ubiquitous at the surface of the moon \citep{lorenz_etal_2008}, Cassini radar measured a bulk dielectric constant 
$\epsilon = 2$ \citep{elachi_etal_2005}, inconsistent with water ice ($\epsilon = 3.1$) or ammonia ice ($\epsilon = 4.5$). The interpretation 
of radar observation is not unique and is still discussed \citep{hofgartner_etal_2020}. Nonetheless, one can assert safely 
that the surface and the near subsurface of Titan are complex systems with horizontal and vertical heterogeneities. This is the reason 
why in some places, a porous deposited organic material a few hundred meters deep may host liquid hydrocarbons, leading to processes like those 
discussed in the present work. In our model of crust, the water-ice thermal conductivity has a value around $5$~W~m$^{-1}$~K$^{-1}$, 
and methane hydrates are much more insulating with values around $0.5$~W~m$^{-1}$~K$^{-1}$, as emphasized by \cite{kalousova_sotin_2020}.
If we adopt bitumen as an analog of Titan's organic sediments, we find a thermal conductivity close to the clathrate one
\citep{nazki_etal_2020}, and, based on a study of Titan's crater relaxation, \cite{schurmeier_dombard_2018} found a thermal conductivity for 
organic-rich sand as low as $0.025$~W~m$^{-1}$~K$^{-1}$.
Hence, the discussion in Sect.~\ref{nonisothermcrust} about Kalousova \& Sotin's results, is also relevant in 
the case of an organics-dominated matrix.
%
%===================================================================================================================================
\begin{figure}[!t]
\begin{center}
\includegraphics[angle=0, width=8.5 cm]{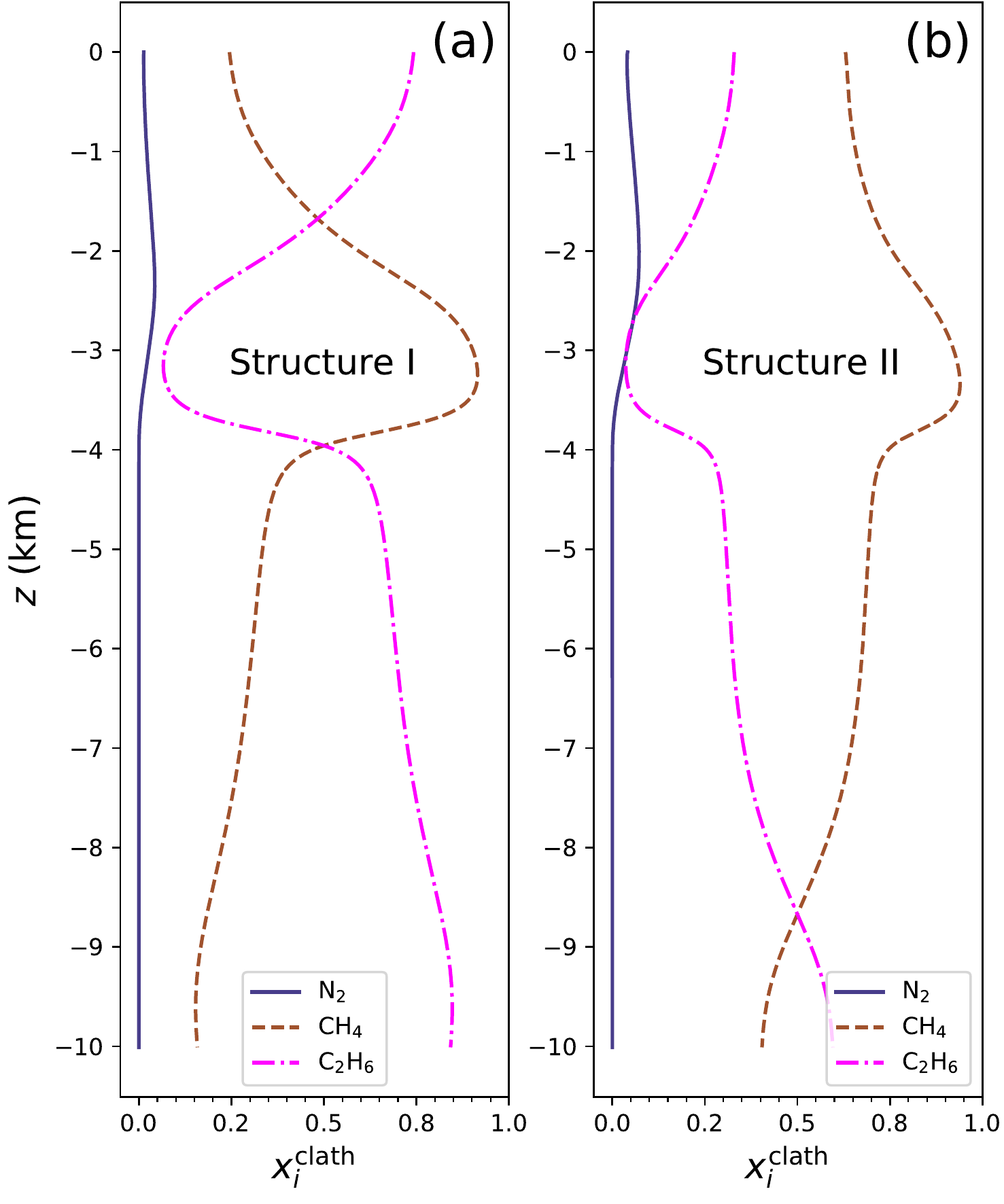}
\caption[]{\label{compClath}Abundances of clathrate hosts (in mole fraction) as a function of the alkanofer depth $z$: 
          (a) for structure I clathrates; (b) for structure II clathrates. In both cases, the assumed liquid composition corresponds
          to the situation depicted in Fig.~\ref{GeoTGrad_crust}(b), for which the crust undergoes a geothermal gradient. This gradient, 
          together with pressure variations, has been taken into account consistently.}
\end{center}
\end{figure}
%===================================================================================================================================
% 
%...................................................................................................................................
\subsection{Liquid-solid interaction}

    Interactions between trapped liquids and solid porous substrates may also be considered. In the case of an organic matrix, if the material 
resembles tholins produced in laboratories, the interaction should be weak since tholins seem to be very poorly soluble in cryogenic 
solvents \citep{carrasco_etal_2009}. The situation is really different when the solid phase is dominated by water, 
which may form clathrate hydrates in the presence of molecules including methane, nitrogen, and ethane. 
This case has been considered many times in previous works 
\citep{lewis_1971,lunine_stevenson_1987,loveday_etal_2001,tobie_etal_2006,choukroun_sotin_2012,kalousova_sotin_2020,vu_etal_2020a}.
The question that arises here is whether an enrichment of N$_2$ can inhibit, in some way, the clathration of CH$_4$ or C$_2$H$_6$.
In search of an answer, we built an equilibrium model for clathrate, based on the \cite{van_der_waals_platteeuw_1959} approach.
The clathration of species contained in the alkanofer liquid is measured by the fractional occupancy for a guest molecule $i$ trapped in
a clathrate crystal forming a cage of size ${\rm S}$ (i.e., small or large), 
with a structure of type ${\rm T}$
(type I orsII). The fractional occupancy is
\begin{equation}
  y_{{\rm S},i}^{\rm T} = \frac{C_{{\rm S}, i}^{\rm T} f_i}{ 1 + \sum_j C_{{\rm S}, j}^{\rm T} f_j}
,\end{equation}
where the $f_i$s are the fugacities of species in the liquid. These quantities have been evaluated according to the equation of state
PC-SAFT already employed in this work. The factors $C_{{\rm S}, i}^{\rm T}$ are the Langmuir constants of guest 
species $i$ in a cage of size ${\rm S}$ and clathrate structure ${\rm T}$.
The expression of $C_{{\rm S}, i}^{\rm T}$ is available in the literature  
\citep[see, for instance,][]{thomas_etal_2008} and 
derived from Kihara potential \citep{sloan_1998}. This theoretical approach enables the calculation of the mole fraction $x_i^{\rm Clath, T}$ of 
guest molecule $i$ in a clathrate of structure type T:
\begin{equation}
x_i^{\rm Clath, T} = \frac{b_{\rm S}^{\rm T} y_{{\rm S}, i}^{\rm T} + b_{\rm L}^{\rm T} y_{{\rm L}, i}^{\rm T}}{
                           b_{\rm S}^{\rm T} \sum_j y_{{\rm S}, j}^{\rm T} + 
                           b_{\rm L}^{\rm T} \sum_j y_{{\rm L}, j}^{\rm T}         }
.\end{equation}
In Fig.~\ref{compClath}, we represent these mole fractions for clathrate structures I and II, taking into account our ``standard''
alkanofer profile and including a geothermal gradient, as depicted in Fig.~\ref{GeoTGrad_crust}(b). As we can see, the presence of N$_2$ in the clathrate cages
remains mostly negligible. Except around $z \sim -3$~km, the hydrocarbons CH$_4$ and C$_2$H$_6$ are clearly the most abundant  
trapped molecules. For structure II, methane is the preponderant guest compound, although we observe an inversion of abundance with ethane at high depth,
meaning deeper than $\sim 9$~km. In the case of a structure I clathrate, ethane is the most abundant trapped
species except at an interval between 
approximately $2000$ and $4000$~m. All these simulations assumed an equilibrium between the clathrate phase and the liquid, the latter being at the 
``diffusion equilibrium''. Implicitly, the existence of a quantity of liquid large enough to fill the parent rock and clathrates, with respect to their porosities, 
is hypothesized. Of course, the straightforward approach described here would deserve a more detailed study.\\
Finally, concerning the interaction solid-liquid, we cannot exclude the existence of more complex, and mainly unknown, physico-chemical processes leading 
to the formation of somewhat exotic species such as the co-crystals recently revealed by laboratory studies 
\citep{vu_etal_2014,vu_etal_2020a,vu_etal_2020b,cable_etal_2014,cable_etal_2019,cable_etal_2020,mcconville_etal_2020,czaplinski_etal_2020}.

%...................................................................................................................................
\subsection{Possible implication for noble gases}

 In 2005, during the descent through Titan's atmosphere, the Gas Chromatograph Mass Spectrometer (GCMS) aboard the 
Huygens probe observed a strong depletion in noble gases \citep{niemann_etal_2005,niemann_etal_2010}.
For instance, the primordial argon isotope, $^{36}$Ar, showed a ratio of $^{36}$Ar/N$_2$ around $10^{-7}$, 
while Kr and Xe were not detected by the GCMS. Given the instrument detection threshold at $10^{-8}$, the atmospheric mole 
fractions of Kr and Xe should be below this upper limit. According to the literature, ratio in numbers Ar/N$_2$, Kr/N$_2,$ and Xe/N$_2$ 
should be around $7\times 10^{-2}$, $5.6 \times 10^{-5}$, and $3.9 \times 10^{-6}$ respectively, if solar values are adopted 
\citep{lodders_2003,lodders_2010,lodders_2019}. Several scenarios have been investigated to explain this apparent lack of rare gases: 
some authors have considered the trapping in possible crustal clathrate layers.
\cite{osegovic_max_2005}, \cite{thomas_etal_2007,thomas_etal_2008}, and \cite{jacovi_bar-nun_2008} have tried to explain the depletion via
a mechanism involving atmospheric haze, while \cite{cordier_etal_2010} studied the dissolution in Titan's liquid phases. None 
of these works brought a fully satisfactory    answer to the problem. 
Here, we can investigate whether noble-gas atoms, which are 
small in size and heavier than N$_2$ or C$_2$H$_6$  (Ar: $39.948$~g~mol$^{-1}$, Kr: $83.798$~g~mol$^{-1}$, Xe: $131.293$~g~mol$^{-1}$),
could accumulate in the deepest alkanofer layers. We performed preliminary computations by replacing N$_2$ from the ternary
mixture and setting both surface mole fractions of the chosen noble gas ``X'' and ethane to $10^{-3}$, which is probably a huge value
for a noble gas but a small one for C$_2$H$_6$. Adding the species X to our initial ternary mixture, N$_2$+C$_2$H$_6$+CH$_4$, would
have required a large and in-depth modification of our model, then forming a quaternary mixture, this is the reason why we simply 
swapped N$_2$ and X. However, this simplified approach allows us to appreciate the global behavior of noble-gas species 
regarding their diffusion in a liquid solvent dominated by methane. As is evident from Table~\ref{DeltaArKrXe}, all the derived vertical enrichment 
ratios (see definition given by Eq.~\ref{vertenrich}) are significantly larger than $1$.
\begin{table}[h]
%\begin{center}
\caption{\label{DeltaArKrXe}Vertical enrichment ratios $\Delta_{\rm X}$ (X= Ar, Kr, Xe) for noble gases.}
%\vspace*{0.5cm}
\begin{tabular}{l|ccc}
                                        & Ar        & Kr        & Xe      \\
\hline
Model with an isothermal crust          & $1.64$    & $11.5$    & $119$   \\
Model including a geothermal gradient   & $1.95$    & $8.98$    & $33.0$  \\ 
\end{tabular}
%\end{center}
\end{table}
Noble gases tend to accumulate at the bottom of the alkanofer, leading to enrichment up to 2 orders 
of magnitude larger than the surface mole fraction fixed to $10^{-3}$ in our test case. The tendency is clear, but 
a quantitative analysis would be required to solve the problem. The starting point of a follow-up study could be the development of an atmosphere-liquid 
phase equilibrium model based on the up-to-date PC-SAFT equation of state.

%...................................................................................................................................
\subsection{Conclusion}

  As stated in the introduction, the fate of ethane produced in the atmosphere poses a puzzling
question. In this work, we addressed the possibility of an ethane storage in the deepest part of an alkanofer, due to the effect of 
molecular diffusion over geological timescales. This kind of scenario has been suggested by the existence of vertical chemical composition 
gradients through terrestrial hydrocarbon reservoirs. Our main hypothesis consists of the existence of an icy matrix, porous enough to 
enable molecular diffusion but with an hydraulic permeability below the threshold at which convective transport starts. 
We found that, even if ethane molecules
are significantly heavier than their methane counterparts, C$_2$H$_6$ does not accumulate at the bottom of the system. In addition, 
and somewhat surprisingly due to its relatively small molecular size, nitrogen seems to easily enrich deep layers. Many important alkanofer 
parameters are unknown; first of all, the composition and the total quantity of available liquids. The geographical and vertical extent of the reservoirs 
are not well constrained; 
the actual porosity, permeability, and connectivity of subsurface reservoirs are also poorly known. 
As a consequence, it is difficult, if not scientifically adventurous, to draw strong conclusions about the mentioned problem. However, arguments 
developed in this work clearly disfavor ethane accumulation in lower parts of subsurface reservoirs, under the action of molecular diffusion. 
Alternatively, if diffusion cannot account for the formation of  significant chemical gradients, other transport processes 
may be more efficient. Several previous works have considered liquid subsurface flows through porous media \citep{hayes_etal_2008,horvath_etal_2016}.
The scale taken into account may be regional \citep[e.g.,][]{horvath_etal_2016,hayes_etal_2017} or global \citep{faulk_etal_2020}; in these
studies, the working fluid is chemically homogenous since it is liquid methane. The present investigations addresses the case of relatively small systems, 
meaning a few kilometers or a few tens of kilometers in size, hydrologically isolated to allow diffusion to be a dominant transport process. 
Future studies may focus on hydrodynamical mixing of local reservoirs with different chemical compositions.\\

   The icy crust of Titan represents an interface between the internal global water ocean and the carbon-rich atmosphere where a plethora 
of organic molecules are produced. From an exobiological perspective, it is of interest to investigate the supply of molecules larger than 
C$_2$-compounds to the underground aqueous environment. The deepest layers of alkanofers should have a low porosity caused by 
compaction, and then in these regions molecular diffusion
should dominate the transport processes of these species. Even if the alkanofers have a maximum depth of a few kilometers, ice convection could 
subsequently bring this carbon-rich material down to the ocean \citep{kalousova_sotin_2020} at a depth of about $80-100$ km. Thus, the 
effect of molecular size on diffusion in liquid CH$_4$+C$_2$H$_6$+N$_2$ mixtures should be explored.\\

 To probe the possible polar Titan alkanofer would require techniques such as deep drilling, resistivity measurements 
\citep{griffiths_barker_1993}, and acoustic or electromagnetic sounding \citep{reynolds_2011}. To date, one mission is planned to perform an in situ
exploration of the surface of Titan: this is the octocopter Dragonfly\footnote{\url{https://dragonfly.jhuapl.edu}} 
\citep{lorenz_etal_2018b}. On board this probe, a scientific package called DraGMet\footnote{Dragonfly Geophysics and Meteorology Package.} 
should contain instruments dedicated to geophysical measurements. However, to the best of our knowledge, the precise capabilities of these 
instruments are not publicly available. It would be surprising if they could probe the Titan subsurface layers down to a depth of several kilometers.
Nonetheless, some particular circumstances could enable the search for clues concerning the behavior and the fate of liquids within underground layers.
For example, Dragonfly will explore a region covered by large dunes and we can imagine a comparison
of liquid composition analysis between samples collected at the base of a dune and at its top.\\

%...................................................................................................................................
\subsection{Data availability}

We make a part of the data related to the present work publicly available on the Open Science platform Zenodo\footnote{\url{https://www.zenodo.org}}
, which belongs to the OpenAIRE project\footnote{\url{https://www.openaire.eu}}. The 
dataset provided allows the reproduction of molecular dynamics simulations used in this work.
Results from GROMACS molecular dynamics simulations performed on two binary mixtures, CH$_4$ (80\%) + C$_2$H$_6$ (20\%) and 
CH$_4$ ($80$\%) + N$_2$($20$\%), and the ternary mixture CH$_4$ ($60$\%) + C$_2$H$_6$($20$\%) + N$_2$ ($20$\%), are available 
online\footnote{\url{https://zenodo.org/record/4975047}} and it is also referenced with the DOI: \url{10.5281/zenodo.4975047}.
Each mixture is composed of $5000$ molecules, and two values of temperature and pressure are considered, namely 
$(T, p) =$ ($90$~K, $1.5$~bar) and ($95$~K, $120$~bar). 
This dataset collects information on trajectories (i.e., atomic positions and velocities, energies, and statistical data) and 
transport properties (diffusion and viscosity) for the six aforementioned cases (three mixtures under two $(T, p)$ conditions).\\
Our PC-SAFT implementation in \texttt{FORTRAN 2008} has been made also publicly available on 
Github\footnote{\url{https://github.com/dcordiercnrs/pcsaft-titan}}
and it has been archived on Zenodo\footnote{\url{https://zenodo.org/record/5085305}} under the DOI: \url{10.5281/zenodo.5085305}.
%
%%%%%%%%%%%%%%%%%%%%%%%%%%%%%%%%%%%%%%%%%%%%%%%%%%%%%%%%%%%%%%%%%%%%%%%%%%%%%%%%%%%%%%%%%%%%%%%%%%%%%%%%%%%%%%%%%%%%%%%%%%%%%%%%%%%%
%%%%%%%%%%%%%%%%%%%%%%%%%%%%%%%%%%%%%%%%%%%%%%%%%%%%%%%%%%%%%%%%%%%%%%%%%%%%%%%%%%%%%%%%%%%%%%%%%%%%%%%%%%%%%%%%%%%%%%%%%%%%%%%%%%%%
%%%%%%%%%%%%%%%%%%%%%%%%%%%%%%%%%%%%%%%%%%%%%%%%%%%%%%%%%%%%%%%%%%%%%%%%%%%%%%%%%%%%%%%%%%%%%%%%%%%%%%%%%%%%%%%%%%%%%%%%%%%%%%%%%%%%
%%%%%%%%%%%%%%%%%%%%%%%%%%%%%%%%%%%%%%%%%%%%%%%%%%%%%%%%%%%%%%%%%%%%%%%%%%%%%%%%%%%%%%%%%%%%%%%%%%%%%%%%%%%%%%%%%%%%%%%%%%%%%%%%%%%%
%%%%%%%%%%%%%%%%%%%%%%%%%%%%%%%%%%%%%%%%%%%%%%%%%%%%%%%%%%%%%%%%%%%%%%%%%%%%%%%%%%%%%%%%%%%%%%%%%%%%%%%%%%%%%%%%%%%%%%%%%%%%%%%%%%%%
%\clearpage

\begin{acknowledgements}
The present research was supported by the Programme National de Plan\'{e}tologie (PNP) of CNRS-INSU co-funded by CNES, and also
partially supported by the HPC center of Champagne-Ardenne (ROMEO) and the French HPC center CINES under project number A0090711976.
D.C. thanks the University of Reims Champagne-Ardenne for granting him with the ``dispositif Mobilité Courte
Entrante et Sortante'' for its stay at the Jet Propulsion Laboratory in 2019.
Part of this work was carried out at the Jet Propulsion Laboratory, California Institute of Technology, 
under a contract with the National Aeronautics and Space Administration.
This work has been done mainly using open-source softwares like \texttt{Python}, \texttt{gfortran}, \texttt{kate}, \texttt{Jupyter Notebook},
\LaTeX{ }under \texttt{GNU/Debian Linux} operating system, and also GROMACS under \texttt{Red Hat Linux}. The authors warmly acknowledge 
the whole Free Software community.
\end{acknowledgements}

%%%%%%%%%%%%%%%%%%%%%%%%%%%%%%%%%%%%%%%%%%%%%%%%%%%%%%%%%%%%%%%%%%%%%%%%%%%%%%%%%%%%%%%%%%%%%%%%%%%%%%%%%%%%%%%%%%%%%%%%%%%%%%%%%%%%
%% Annexe :
\appendix
% --------------------------------------------------------------------------------------------
\section{List of variables}
\label{VariableList}

\begin{itemize}
   \item[\textbullet] $P$ and $T$: local pressure (Pa) and temperature (K).\\
   
   \item[\textbullet] $z$: depth below Titan's surface (m).\\
   
   \item[\textbullet] $x_i$: local mole fraction of compound $i$.\\
   
   \item[\textbullet] $M_i$: molecular mass of species $i$, $\bar{M}$ mean molecular mass of considered material (kg~mol$^{-1}$).\\
   
   \item[\textbullet] $\bar{V}_i$: partial molar volume of component $i$ (m$^3$~mol$^{-1}$).\\
   
   \item[\textbullet] $f_i$: fugacity of component $i$ (Pa).\\
   
   \item[\textbullet] $\mathcal{D}_{ij}$: fickean molecular diffusion coefficient (m$^2$~s$^{-1}$) of compound $i$ in species $j$.\\
   
   \item[\textbullet] $D^M$ (kg~mol$^{-1}$~m$^2$~s$^{-1}$), $D^P$ (kg~mol$^{-1}$~m$^2$~s$^{-1}$~Pa$^{-1}$) and $D^T$
        (kg~mol$^{-1}$~m$^2$~s$^{-1}$~K$^{-1}$): tensors introduced by \cite{ghorayeb_firoozabadi_2000} are not to be confused
        with the $\mathcal{D}_{ij}$'s.\\
        
   \item[\textbullet] $g$, $g_{\rm Tit}$, $g_{\rm Earth}$: gravity (generic notation, Titan's value, Earth's value, m~s$^{-2}$).\\
   
   \item[\textbullet] $k_{ij,T}$: thermal diffusion ratio (dimensionless).\\
   
   \item[\textbullet] $\alpha_{ij,T}$: thermal diffusion coefficient, $k_{ij,T}= x_i x_j \alpha_{ij, T}$ (dimensionless).\\
   
   \item[\textbullet] $\rho_{\rm eff}$: effective density (kg~m$^{-3}$) of the alkanofer material,
        $\rho_{\rm eff}= \Pi \, \rho_{\rm liq} + (1-\Pi) \, \rho_{\rm ice,}$ with $\Pi$ being the porosity, $\rho_{\rm liq}$ the density
        of the liquid (kg~m$^{-3}$), and $\rho_{\rm ice}$ the density of the icy matrix (kg~m$^{-3}$).\\
        
   \item[\textbullet] $\mu_i$: chemical potential of component $i$ (J~mol$^{-1}$), $\mu^{res}_i$ is the residual chemical potential of 
        component $i$ (J~mol$^{-1}$) the term ``residual'' relates to a difference between the physical quantities for the actual fluid 
        and the corresponding ideal gas.\\
   
   \item[\textbullet] $h_i$: residual molar enthalpy of component $i$ (J~mol$^{-1}$).\\
   
   \item[\textbullet] $\bar{h}_i$: residual partial molar enthalpy of component $i$ (J~mol$^{-1}$).\\
   
   \item[\textbullet] $\tilde{a}^{res}$: residual and reduced (i.e., molar) Helmholtz free energy (J~mol$^{-1}$).\\

   \item[\textbullet] $Z$: compressibility factor of the considered fluid (dimensionless).\\
   
   \item[\textbullet] $K_{p,sc}$: permeability (m 2) of the porous media that connects two reservoirs.\\
   
   \item[\textbullet] $\eta$: dynamic viscosity (Pa~s); $\nu$ is the kinematic viscosity ($\eta=\rho \nu$).\\
   
   \item[\textbullet] $\Delta_i$: vertical enrichment ratios (dimensionless) for species $i$.
\end{itemize}

% --------------------------------------------------------------------------------------------
\section{Model of molecular, pressure, and thermal diffusion in a nonideal ternary mixture}
\label{ModelDiffTernary}

    Our model is essentially based on the formalism described in \cite{ghorayeb_firoozabadi_2000} for nonideal multicomponent
mixtures. In this work, for the purpose of clarity we mainly kept the notation adopted in the mentioned reference. Thus, the 
derivative of the logarithm of the fugacity $f_i$ of component $i$, with respect to the mole fraction $x_j$ of the species $j$, is
\begin{equation}
   f_{ij} = \left(\frac{\partial \mathrm{ln} \, f_i}{\partial x_j}\right)_{x_{i\ne j}, T, P}
.\end{equation}
The fugacity $f_i$ is evaluated according to the PC-SAFT equation of state, already used in many Titan-related papers 
\citep[e.g.,][]{tan_etal_2013,cordier_etal_2017a}. Generally speaking, the fugacity may be written as $f_i = \Phi_i x_i P$, where 
$\Phi_i$ is the fugacity coefficient directly provided by PC-SAFT, and $P$ stands for the pressure. The derivatives are obtained by
finite differences. Employing a very usual
notation, $M_i$ represents the molecular mass of species $i$, while the average 
molecular mass is noted $\bar{M}$. In the following, we also use the reduced expression
\begin{equation}
   a_{ij} = \frac{M_i M_j}{\bar{M}}
.\end{equation}
In a usual way, the fickean molecular diffusion coefficients of species $i$ in component $j$ are denoted $\mathcal{D}_{ij}$ (m$^{2}$~s$^{-1}$).
The gas constant is simply written as $R$.
Four constants are introduced \citep[see Eqs.~36--39, in][]{ghorayeb_firoozabadi_2000}:
\begin{equation}
   c_1 = \frac{M_1 x_1 + M_3 x_3}{M_1} f_{12} + x_2 f_{22}
,\end{equation}
\begin{equation}
   c_2 = x_1 f_{12} + \frac{M_2 x_2 + M_3 x_3}{M_2} f_{22}
,\end{equation}
\begin{equation}
   c_3 = \frac{M_1 x_1 + M_3 x_3}{M_1} f_{11} + x_3 f_{21}
,\end{equation}
\begin{equation}
   c_4 = x_1 f_{11} + \frac{M_2 x_2 + M_3 x_3}{M_2} f_{21}
.\end{equation}
In addition, we define
\begin{equation}
   c_5 = \frac{M_1 x_1 + M_3 x_3}{M_1} f_{12} + x_2 f_{21}
.\end{equation}
The elements of the molecular diffusion matrix $D^M$ are given by \citep[see Eqs.~40--43, in][]{ghorayeb_firoozabadi_2000}
\begin{equation}
   D_{11}^M = a_{13} \mathcal{D}_{13} M_1 x_1 \, \left(c_3 + c_4 \frac{L_{12}}{L_{11}}\right)
,\end{equation}
\begin{equation}
   D_{12}^M = a_{13} \mathcal{D}_{13} M_1 x_1 \, \left(c_1 + c_2 \frac{L_{12}}{L_{11}}\right)
,\end{equation}
\begin{equation}
   D_{21}^M = a_{23} \mathcal{D}_{23} M_2 x_2 \, \left(c_4 + c_3 \frac{L_{21}}{L_{22}}\right)
,\end{equation}
\begin{equation}
   D_{22}^M = a_{23} \mathcal{D}_{23} M_2 x_2 \, \left(c_2 + c_1 \frac{L_{21}}{L_{22}}\right)
.\end{equation}
  The $L_{ij}$s are the so-called Onsager phenomenological coefficients given by
\begin{equation}\label{L11}
   L_{11} = \frac{c M_3 x_3 a_{13} \mathcal{D}_{13} M_1 x_1}{R} 
,\end{equation}
\begin{equation}\label{L22}
   L_{22} = \frac{c M_3 x_3 a_{23} \mathcal{D}_{23} M_2 x_2}{R}
.\end{equation}
Using Eq.~(49) of \cite{ghorayeb_firoozabadi_2000} for $i=1$ and $j=2$, we can derive
\begin{equation}\label{L12}
   L_{12} = L_{21} = \frac{a_{13} \mathcal{D}_{13} M_1 x_1 (c_1c_3 - c_1c_5)}{\frac{a_{23} \mathcal{D}_{23} M_2 x_2 (c_1c_4-c_2c_3)}{L_{22}} 
                                                                  - \frac{a_{13} \mathcal{D}_{13} M_1 x_1 (c_1c_4 - c_2c_5)}{L_{11}}}  
.\end{equation}
  The terms related to the barodiffusion are provided by Eqs.~(46) and (47) of  \cite{ghorayeb_firoozabadi_2000}:
\begin{equation}\label{D1_P}
\begin{split}
   D_1^P =  a_{13} \mathcal{D}_{13} \frac{M_1 x_1}{RT} \, & \left\{ \frac{M_1 x_1 + M_3 x_3}{M_1} \bar{V}_1 + x_2 \bar{V}_2 - \frac{1}{c} + \right. \\
           &        \left.\left(\frac{M_2 x_2 + M_3 x_3}{M_2} \bar{V}_2 + x_1 \bar{V}_1 - \frac{1}{c}\right) \frac{L_{12}}{L_{11}} \right\}
\end{split}
,\end{equation}
\begin{equation}\label{D2_P}
\begin{split}
   D_2^P = a_{23} \mathcal{D}_{23} \frac{M_2 x_2}{RT} & \left\{ \frac{M_2 x_2 + M_3 x_3}{M_2} \bar{V}_2 + x_1 \bar{V}_1 - \frac{1}{c} + \right. \\
          &        \left.\left(\frac{M_1 x_1 + M_3 x_3}{M_1} \bar{V}_1 + x_2 \bar{V}_2 - \frac{1}{c}\right) \frac{L_{21}}{L_{22}}\right\}
\end{split}
.\end{equation}
Finally, concerning the thermal diffusion, we have  \citep[see Eqs.~44 and 45 in][]{ghorayeb_firoozabadi_2000}:
\begin{equation}\label{D1_T}
   D_1^T = a_{13} \mathcal{D}_{13} \bar{M} \frac{k_{T\,13}}{T}
,\end{equation}
\begin{equation}\label{D2_T}
   D_2^T = a_{23} \mathcal{D}_{23} \bar{M} \frac{k_{T\,23}}{T}
.\end{equation}
The thermal diffusion ratio $k_{T \, i,j}$ (dimensionless) is a function of $\alpha_{T\, i,j}$, the thermal diffusion coefficient (dimensionless), 
according to the formula $k_{T \, i,j}= \alpha_{T\, i,j} x_i \, x_j$.
   In all these equations, the density of the liquid $\rho$ (kg~m$^{-3}$), used in $c= \rho/\bar{M}$, the partial molar volume $\bar{V}_i$ 
(m$^3$~mol$^{-1}$), the fugacities $f_i,$ and the thermal diffusion coefficient $ \alpha_{T\, i,j}$ are derived from the PC-SAFT equation of state.
The equation that provides $\bar{V}_i$ has been established in a previous work \citep[see Eq.~19 in][]{cordier_etal_2019}.
    The thermal diffusion coefficient $\alpha_{ij,T}$ is derived from PC-SAFT thanks to the Haase formula \citep{haase_1969,pan_etal_2006}:
\begin{equation}\label{alphaHaase}
  \alpha^{\rm Haase}_{ij, T} = \frac{RT}{x_i \left(\frac{\partial\mu_i}{\partial x_i}\right)} \, 
     \left\{\alpha^0_{ij,T} + \frac{M_i \frac{\bar{h}_j^{res}}{RT} - M_j \frac{\bar{h}_i^{res}}{RT}}{M_i x_i + M_j x_j}\right\}
,\end{equation}
where the $M_i$s and $x_i$s are, respectively, the molecular weights and the mole fractions, $\mu_i$ is the chemical potential of component $i$,
$\alpha^0_{ij,T}$ represents the thermal diffusion coefficient for the corresponding ideal gas of the couple of species $(i,j)$, and 
$\bar{h}_i^{res}$ is the residual partial molar enthalpy of species $i$, also provided by PC-SAFT. 

% --------------------------------------------------------------------------
\section{Structure of the crust}
\label{structcrust}

Following the Titan interior model published by \cite{sohl_etal_2014}, the properties of the icy crust may be retrieved by 
integrating
\begin{equation}\label{eqCrust_hydro}
  \frac{\mathrm{d}P}{\mathrm{d}r} = -\rho g_{\rm Tit} 
,\end{equation}
\begin{equation}\label{eqCrust_energy}
  \frac{\mathrm{d} q}{\mathrm{d} r} = -2\frac{q}{r}
,\end{equation}
\begin{equation}\label{eqCrust_heatFlux}
   \frac{\mathrm{d} T}{\mathrm{d} r} = -\frac{q}{N_{u_r} k_{\rm c}}
,\end{equation}
where $P$, $T,$ and $r$ are, respectively, the pressure (Pa), the temperature $({\rm K}),$ and the distance from Titan's center of mass (m). The density of the 
medium (kg~m$^{-3}$) is denoted $\rho$ as usual, $k_{\rm c}$ represents the thermal conductivity (W~m$^{-1}$~K$^{-1}$) of the material, while
$N_{u_r}$ is the local Nusselt number, which parameterizes the strength of the convective heat flux. $N_{u_r}$ is 
given by \citep[see Eq.~7 in ][]{sohl_etal_2014}
\begin{equation}
   N_{u_r} = \left(1 + \frac{k_{\rm v}}{k_{\rm c}}\right) \, 
             \left(\frac{q}{q - k_{\rm v} \, \left(\frac{\mathrm{d}T}{\mathrm{d}r}\right)_{\rm ad}}\right)
,\end{equation}
where the ``adiabatic gradient'' $(\mathrm{d}T/\mathrm{d}r)_{\rm ad}$ is provided by
\begin{equation}\label{dTsdr_ad}
   \left(\frac{\mathrm{d}T}{\mathrm{d}r}\right)_{\rm ad} = -\frac{g_{\rm Tit} \, \alpha \, T}{C_{\rm P}}
,\end{equation}
and the ``conductive gradient'' may be defined as
\begin{equation}
   \left(\frac{\mathrm{d}T}{\mathrm{d}r}\right)_{\rm cond} = -\frac{q}{k_{\rm c}}
.\end{equation}
According to the mixing length theory \citep{kamata_2018}, if
\begin{equation}
   \left(\frac{\mathrm{d}T}{\mathrm{d}r}\right)_{\rm cond} \le \left(\frac{\mathrm{d}T}{\mathrm{d}r}\right)_{\rm ad}
,\end{equation}
then
\begin{equation}\label{kv}
  k_{\rm v} = \frac{\alpha \, C_{\rm P} \, \rho^2 \, g_{\rm Tit} \, l^4}{18 \, \eta} \,
              \left[\left(\frac{\mathrm{d}T}{\mathrm{d}r}\right)_{\rm ad} - \left(\frac{\mathrm{d}T}{\mathrm{d}r}\right)_{\rm cond}\right]
;\end{equation}
otherwise, $k_{\rm v} = 0$. In Eqs.~\ref{dTsdr_ad} and~\ref{kv}, $\alpha$, $C_{\rm P,}$ and $\eta$ are, respectively, the thermal expansion coefficient 
(K$^{-1}$) of water-ice I$_{\rm h}$, the specific heat capacity (J~kg$^{-1}$~K$^{-1}$), and the viscosity (Pa~s). The parameter $l$ has been taken equal to 
the local pressure scale height $H_{\rm P}$ which is evaluated by
\begin{equation}\label{Hp}
  H_{\rm P} = -\frac{1}{P} \,  \frac{\mathrm{d}P}{\mathrm{d}r}
.\end{equation}
The viscosity of the crust $\eta$ is essentially unknown. \cite{sohl_etal_2014} chose a value ($10^{20}$~Pa~s) large enough to impede 
solid convection, since arguments have been put forward concerning the absence of convection in Titan's crust \citep{nimmo_bills_2010}.
The system of ordinary differential
equations formed by Eqs.~\ref{eqCrust_hydro}, \ref{eqCrust_energy}, and \ref{eqCrust_heatFlux} can easily be solved by a Runge-Kutta method, 
taking the values at the surface as boundary conditions: $P=P_s= 1.5$~bar, $q= q_s= 3.11 \times 10^{-3}$~W~m$^{-2,}$ 
and $T=T_s= 94$~K \citep{sohl_etal_2014}.

%\clearpage
\bibliographystyle{aa}
%\bibliography{../BIB/bibliographie_planeto_2020,../BIB/bibliographie_CompGradient_2020,../BIB/bibliographie_bubbles_2020,../BIB/bibliographie_CriticalPoint_2020,../BIB/bibliographie_Divers_2020, ../BIB/bibliographie_Hydrologie_2020,../BIB/bibliographie_volcanism_2020}
\def\sciam{Sci.
  Am.}\def\nature{Nature}\def\nat{Nature}\def\science{Science}\def\natastro{Nat.
  Astron.}\def\natgeo{Nat. Geosci.}\def\natcom{Nat.
  Commun.}\def\pnas{PNAS}\def\AnnderPhys{‎Ann. Phys.
  (Berl.)}\def\icarus{Icarus}\def\pss{Planet. Space
  Sci.}\def\psj{PSJ}\def\planss{Planet. Space Sci.}\def\ssr{Space Sci.
  Rev.}\def\solsr{Sol. Syst. Res.}\def\ApSSP{ApSSP}\def\psj{Planet. Sci.
  J.}\def\jqsrt{J. Quant. Spectrosc. Radiat. Transfer}\def\expastro{Exp.
  Astron.}\def\jcis{‎J. Colloid Interface
  Sci.}\def\aap{A\&A}\def\apj{ApJ}\def\apjl{ApJL}\def\apjs{ApJS}\def\aj{AJ}\def\mnras{MNRAS}\def\araa{Annu.
  Rev. Astron. Astrophys.}\def\pasj{Publ. Astron. Soc.
  Jpn.}\def\apss{Astrophys. Space Sci.}\def\pasp{Publ. Astron. Soc.
  Pac.}\def\expastron{Exp. Astron.}\def\asr{Adv. Space
  Res.}\def\galaxies{Galaxies}\def\astrobiol{Astrobiology}\def\areps{Annu. Rev.
  Earth Planet. Sci.}\def\georl{Geophys. Res. Lett.}\def\grl{Geophys. Res.
  Lett.}\def\jgr{J. Geophys. Res.}\def\gca{Geochim. Cosmochim.
  Ac.}\def\epsl{Earth Planet. Sci. Lett.}\def\ess{Earth Space
  Sci.}\def\plasci{Planet. Sci.}\def\ggg{Geochem. Geophys.
  Geosyst.}\def\rmg{Rev. Mineral. Geochem.}\def\gji{Geophys. J.
  Int.}\def\tpm{Transport Porous Med.}\def\philtrans{Phil.
  Trans.}\def\faradis{Farad. Discuss.}\def\jcis{‎J. Colloid Interface
  Sci.}\def\jfm{J. Fluid Mech.}\def\physflu{Phys. Fluids}\def\pachem{Pure Appl.
  Chem.}\def\jpcA{J. Phys. Chem. A}\def\chemrev{Chem. Rev.}\def\AppOpt{Appl.
  Opt.}\def\nature{Nature}\def\nat{Nature}\def\science{Science}\def\natastro{Nat.
  Astron.}\def\natphys{Nat. Phys.}\def\natgeo{Nat. Geosci.}\def\natcom{Nat.
  Commun.}\def\scirep{Sci. Rep.}\def\science{Sci}\def\jced{J. Chem. Eng.
  Data}\def\fpe{Fluid Phase Equilibria}\def\iecr{Ind. Eng. Chem.
  Res.}\def\aichej{AIChE J.}\def\pt{Powder Technol.}\def\etfs{Exp. Therm. Fluid
  Sci.}\def\jgr{J. Geophys. Res.}\def\gca{Geochim. Cosmochim. Acta}\def\jcp{J.
  Chem. Phys.}\def\jpcl{J. Phys. Chem. Lett.}\def\jcis{‎J. Colloid Interface
  Sci.}\def\jcsft{J. Chem. Soc. Faraday Trans.}\def\jpcB{J. Phys. Chem.
  B}\def\jsf{J. Supercrit. Fluids}\def\enerp{Energy Procedia}\def\aichej{AlChE
  J.}\def\IECPDD{Ind. Eng. Chem. Process Des. Dev.}\def\pre{Phys. Rev.
  E.}\def\orggeoch{Org.
  Geochem.}\def\nature{Nature}\def\nat{Nature}\def\science{Sci}\def\pnas{Proc.
  Natl. Acad. Sci. U.S.A.}\def\jced{J. Chem. Eng. Data}\def\fpe{Fluid Phase
  Equilibria}\def\iecr{Ind. Eng. Chem. Res.}\def\aichej{AIChE J.}\def\pt{Powder
  Technol.}\def\etfs{Exp. Therm. Fluid Sci.}\def\mineng{Miner.
  Eng.}\def\intjminprocess{Int. J. Miner. Process}\def\jcolintersci{J. Colloid
  Interface
  Sci.}\def\nature{Nature}\def\nat{Nature}\def\science{Science}\def\natastro{Nat.
  Astron.}\def\natphys{Nat. Phys.}\def\natgeo{Nat. Geosci.}\def\natcom{Nat.
  Commun.}\def\scirep{Sci. Rep.}\def\science{Sci}\def\jced{J. Chem. Eng.
  Data}\def\fpe{Fluid Phase Equilibria}\def\iecr{Ind. Eng. Chem.
  Res.}\def\aichej{AIChE J.}\def\pt{Powder Technol.}\def\etfs{Exp. Therm. Fluid
  Sci.}\def\jgr{J. Geophys. Res.}\def\gca{Geochim. Cosmochim. Acta }\def\jcp{J.
  Chem. Phys.}\def\jpcl{J. Phys. Chem. Lett.}\def\jcis{‎J. Colloid Interface
  Sci.}\def\jcsft{J. Chem. Soc. Faraday Trans.}\def\jpcB{J. Phys. Chem.
  B}\def\jsf{J. Supercrit. Fluids}\def\enerp{Energy Procedia}\def\aichej{AlChE
  J.}\def\IECPDD{Ind. Eng. Chem. Process Des. Dev.}\def\pre{Phys. Rev.
  E.}\def\prl{Phys. Rev.
  Lett.}\def\nature{Nature}\def\nat{Nature}\def\science{Science}\def\natastro{Nat.
  Astron.}\def\natgeo{Nat. Geosci.}\def\natcom{Nat. Commun.}\def\scirep{Sci.
  Rep.}\def\sciad{Sci.
  Adv.}\def\aap{A\&A}\def\apj{ApJ}\def\apjl{ApJL}\def\apjs{ApJS}\def\aj{AJ}\def\mnras{MNRAS}\def\araa{Annu.
  Rev. Astron. Astrophys.}\def\pasj{Publ. Astron. Soc.
  Jpn.}\def\apss{Astrophys. Space Sci.}\def\pasp{Publ. Astron. Soc.
  Pac.}\def\expastron{Exp. Astron.}\def\asr{Adv. Space
  Res.}\def\galaxies{Galaxies}\def\science{Sci}\def\jced{J. Chem. Eng.
  Data}\def\fpe{Fluid Phase Equilibria}\def\iecr{Ind. Eng. Chem.
  Res.}\def\aichej{AIChE J.}\def\pt{Powder Technol.}\def\etfs{Exp. Therm. Fluid
  Sci.}\def\jgr{J. Geophys. Res.}\def\gca{Geochim. Cosmochim.
  Acta}\def\chemgeol{Chem Geol.}\def\jcp{J. Chem. Phys.}\def\jcis{‎J. Colloid
  Interface Sci.}\def\jcsft{J. Chem. Soc. Faraday Trans.}\def\jpcB{J. Phys.
  Chem. B}\def\jsf{J. Supercrit. Fluids}\def\enerp{Energy
  Procedia}\def\aichej{AlChE J.}\def\IECPDD{Ind. Eng. Chem. Process Des.
  Dev.}\def\EF{Energy Fuels}\def\jacs{J. Am. Chem.
  Soc.}\def\nature{Nature}\def\nat{Nature}\def\science{Science}\def\natastro{Nat.
  Astron.}\def\natphys{Nat. Phys.}\def\natgeo{Nat. Geosci.}\def\natcom{Nat.
  Commun.}\def\scirep{Sci. Rep.}\def\science{Sci}\def\jced{J. Chem. Eng.
  Data}\def\fpe{Fluid Phase Equilibria}\def\iecr{Ind. Eng. Chem.
  Res.}\def\aichej{AIChE J.}\def\pt{Powder Technol.}\def\etfs{Exp. Therm. Fluid
  Sci.}\def\jgr{J. Geophys. Res.}\def\gca{Geochim. Cosmochim. Acta}\def\jcp{J.
  Chem. Phys.}\def\jpcl{J. Phys. Chem. Lett.}\def\jcis{‎J. Colloid Interface
  Sci.}\def\jcsft{J. Chem. Soc. Faraday Trans.}\def\jpcB{J. Phys. Chem.
  B}\def\jsf{J. Supercrit. Fluids}\def\enerp{Energy Procedia}\def\aichej{AlChE
  J.}\def\IECPDD{Ind. Eng. Chem. Process Des. Dev.}\def\pre{Phys. Rev.
  E.}\def\orggeoch{Org. Geochem.}\def\sciam{Sci.
  Am.}\def\nature{Nature}\def\nat{Nature}\def\science{Science}\def\natastro{Nat.
  Astron.}\def\natgeo{Nat. Geosci.}\def\natcom{Nat.
  Commun.}\def\pnas{PNAS}\def\AnnderPhys{‎Ann. Phys.
  (Berl.)}\def\icarus{Icarus}\def\pss{Planet. Space Sci.}\def\planss{Planet.
  Space Sci.}\def\ssr{Space Sci. Rev.}\def\solsr{Sol. Syst.
  Res.}\def\expastro{Exp. Astron.}\def\jcis{‎J. Colloid Interface
  Sci.}\def\aap{A\&A}\def\apj{ApJ}\def\apjl{ApJL}\def\apjs{ApJS}\def\aj{AJ}\def\mnras{MNRAS}\def\araa{Annu.
  Rev. Astron. Astrophys.}\def\pasj{Publ. Astron. Soc.
  Jpn.}\def\apss{Astrophys. Space Sci.}\def\pasp{Publ. Astron. Soc.
  Pac.}\def\expastron{Exp. Astron.}\def\asr{Adv. Space
  Res.}\def\astrobiol{Astrobiology}\def\areps{Annu. Rev. Earth Planet.
  Sci.}\def\georl{Geophys. Res. Lett.}\def\jgr{J. Geophys.
  Res.}\def\gca{Geochim. Cosmochim. Ac.}\def\epsl{Earth Planet. Sci.
  Lett.}\def\plasci{Planet. Sci.}\def\ggg{Geochem. Geophys.
  Geosyst.}\def\rmg{Rev. Mineral. Geochem.}\def\tpm{Transport Porous
  Med.}\def\philtrans{Phil. Trans.}\def\faradis{Farad. Discuss.}\def\jcis{J.
  Colloid Interface Sci.}\def\jfm{J. Fluid Mech.}\def\physflu{Phys.
  Fluids}\def\pachem{Pure Appl. Chem.}\def\jpcA{J. Phys. Chem.
  A}\def\chemrev{Chem. Rev.}

%\end{linenumbers}

%%%%%%%%%%%%%%%%%%%%%%%%%%%%%%%%%%%%%%%%%%%%%%%%%%%%%%%%%%%%%%%%%%%%%%%%%%%%%%%%%%%%%%%%%%%%%%%%%%%%%%%%%%%%%%%%%%%%%%%%%%%%%%%%%%%%
%% 
\end{document}